\documentclass[a4paper,12pt]{article}
\usepackage[a4paper,text={16.8cm,22.4cm}]{geometry}
\usepackage{amsmath,braket,bm}
\usepackage{cite}
\usepackage{graphicx}
\usepackage{hepunits}
\usepackage{color}
\usepackage{multirow}
\usepackage{amssymb}
\usepackage[utf8]{inputenc}
\usepackage{float}
\usepackage{hyperref}

\allowdisplaybreaks[4]

\begin{document}

\begin{titlepage}

\begin{center}

\textbf{ \Large Towards a new approximation for pair-production and associated-production of the Higgs boson }

\vspace{7ex}

\textsc{ Xiaofeng Xu$^{a}$, Li Lin Yang$^{a,b,c}$ }

\vspace{2ex}

\textsl{
${}^a$School of Physics and State Key Laboratory of Nuclear Physics and Technology,\\
Peking University, Beijing 100871, China
\\[0.3cm]
${}^b$Collaborative Innovation Center of Quantum Matter, Beijing, China
\\[0.3cm]
${}^c$Center for High Energy Physics, Peking University, Beijing 100871, China
\\[0.3cm]
}
\end{center}

\vspace{4ex}

\begin{abstract}

We propose that loop integrals with internal heavy particles can be evaluated by expanding in the limit of small external masses. This provides a systematically improvable approximation to the integrals in the entire phase space, and works particularly well for the high energy tails of kinematic distributions (where the usual $1/M$ expansions cease to be valid). We demonstrate our method using Higgs boson pair production as an example. We find that at both one-loop and two-loop, our method provides good approximations to the integrals appearing in the scattering amplitudes. Comparing to existing expansion methods, our method are not restricted to a special phase space region. Combining our efficient method to compute the two-loop amplitude with an infrared subtraction method for the real emission corrections, we expect to have a fast and reliable tool to calculate the differential cross sections for Higgs boson pair production. This will be useful for phenomenological studies and for the extraction of the Higgs self-coupling from future experimental data. Our method can also be applied to other processes, such as the associated production of the Higgs boson with a jet or a $Z$ boson.

\end{abstract}

\end{titlepage}

\section{Introduction}

Higgs boson pair production and Higgs boson production associated with a jet ($H + j$) are both important processes at the Large Hadron Collider (LHC). Higgs boson pair production can be used to measure the trilinear self-coupling of the Higgs boson, which is essential to understand the electroweak symmetry breaking in the Standard Model (SM). It also happens that in the SM, contributions from different diagrams to the Higgs pair production cancel each other delicately, leading to a rather small production cross section \cite{Glover:1987nx, Plehn:1996wb}. In a new physics model beyond the SM, such cancellations are not generically expected, and the new physics contribution can potentially be much larger than the SM one. This will lead to large deviations in the total rate as well as kinematic distributions, which have been studied in many new physics models (see, e.g., \cite{deFlorian:2016spz} and references therein) and in the effective field theory framework \cite{Goertz:2014qta, Azatov:2015oxa}. It is expected that Higgs boson pairs will be detected during the high luminosity phase of the LHC, or even earlier if new physics effects enhance the production rate. This will give us hints about the nature of electroweak phase transition in and beyond the SM.

The $H + j$ production process is also useful both for testing the SM and for probing new physics. The recoil of the additional jet gives rise to a non-zero transverse momentum ($p_T$) of the Higgs boson, which is a highly important observable actively being measured by the ATLAS \cite{Aad:2014lwa, Aaboud:2018ezd} and the CMS \cite{Khachatryan:2015rxa, CMS:2018hhg} experiments at the LHC. The high energy tail of the $p_T$ distribution will allow us to probe possible loop corrections from heavy particles which may not be visible in the single Higgs boson production rate.

\begin{figure}[h!]
\centering
\includegraphics[width=0.7\textwidth]{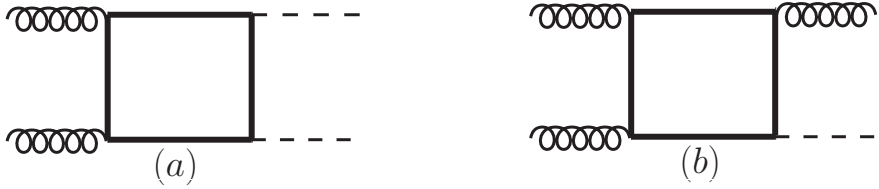}
\caption{Typical Feynman diagrams for (a) Higgs boson pair production and (b) Higgs boson production associated with a jet in the gluon-fusion channel at the leading order.}
\label{fig:LO}
\end{figure}

Due to the importance of these processes, lots of theoretical efforts have been devoted to their precision predictions. At the LHC, the main production mechanism for Higgs boson pairs and for $H + j$ is gluon fusion, in which the Higgs bosons couple to gluons via a top-quark loop. Typical Feynman diagrams at the leading order (LO) are depicted in Figure~\ref{fig:LO}. The LO contributions for both processes have already been known for 30 years, \cite{Glover:1987nx, Plehn:1996wb} for Higgs boson pair and \cite{Ellis:1987xu, Baur:1989cm} for $H+j$. However, the calculations beyond the LO in quantum chromodynamics (QCD) turned out to be highly complicated, due to the multiple energy scales involved in these processes. This makes the higher order amplitudes rather lengthy, and also forbids an analytic representation of the multi-loop integrals in terms of commonly known mathematical functions. In \cite{Borowka:2016ehy, Borowka:2016ypz}, the exact NLO QCD corrections to Higgs boson pair production were finally calculated, using a fully numerical method to evaluate the difficult two-loop integrals. The exact NLO QCD corrections to $H + j$ production were also calculated with a similar method in \cite{Jones:2018hbb}. The numerical nature of these calculations make them rather time-consuming and not very flexible to perform comprehensive phenomenological studies.

Because of the difficulties in the evaluation of multi-loop integrals with many scales, various approximation methods in certain kinematic limits have been proposed to study production processes of the Higgs boson. A common goal of these methods is to reduce the number of scales involved in the loop integrands, such that the resulting integrals are simpler to calculate. The earliest and easiest one is the Higgs effective field theory (HEFT) approach. In this approach, one takes the top quark $m_t$ to infinity, such that the top quark is integrated out leading to effective operators involving Higgs fields and gluon fields. In this way, the NLO QCD corrections only involve one-loop integrals (compared to two-loop ones in the exact calculation) and the scale $m_t$ does not appear in propagators anymore. These greatly simplify the higher order calculations. In fact, the single Higgs boson production cross section has been calculated in this approach to the next-to-next-to-next-to-leading order (NNNLO) in QCD \cite{Anastasiou:2015ema}. In \cite{Boughezal:2013uia, Chen:2014gva, Boughezal:2015dra, Boughezal:2015aha, Caola:2015wna, Chen:2016zka}, the $H+j$ process were calculated at the next-to-next-to-leading order (NNLO) in the HEFT way. As for the Higgs boson pair production, the NNLO corrections in the infinite $m_t$ limit were calculated in \cite{deFlorian:2013uza, deFlorian:2013jea, deFlorian:2016uhr}, and also the third order virtual corrections are available \cite{Baikov:2009bg, Gehrmann:2010ue, Banerjee:2018lfq}.

The HEFT approach provides reasonable approximations for the total cross section, as well as for differential cross sections in the kinematic regions where the Higgs bosons are not highly boosted. However, this approach has some drawbacks. First, with the increasing of the perturbative order, the $1/m_t$ power corrections to the effective theory becomes non-negligible. This is the case for the inclusive single Higgs boson production, where the finite $m_t$ effect at the NLO \cite{Spira:1995rr} is similar in size to the NNNLO contribution \cite{Anastasiou:2015ema}, and has to be taken into account for phenomenology. More severely, if one considers the kinematic distributions, the HEFT is simply not valid when the energy of the Higgs boson is comparable to or larger than the top quark mass. These high energy tails of the differential distributions, on the other hand, are sensitive to new physics effects and are phenomenologically much more interesting. Therefore, the HEFT approach requires some refinements in these two aspects. 

One improvement to the HEFT approximation is performing the $1/m_t$ expansion of the loop integrals to higher powers. This corresponds to including higher dimensional operators in the effective field theory. In this approach, one takes the limit where $m_t$ is much larger than the energies of external particles, and performs a power expansion in terms of $p^\mu/m_t$ for the loop integrand, where $p^\mu$ is some external momentum. The remaining integrals only involves one mass scale, and are very easy to evaluate. This has been done at the NLO for $H + j$ production \cite{Neumann:2016dny, Neumann:2018bsx} and for Higgs boson pair production \cite{Grigo:2013rya, Grigo:2015dia, Degrassi:2016vss}. It was found that the expansion converges rather well for the total cross sections as well as in the low energy regions of differential distributions. For these observables, the $1/m_t$ expansion therefore provides a fast method to obtain predictions with enough precision.

When the energies of external particles increase above the top quark mass, however, the $1/m_t$ expansion quickly fails to converge. And one does not expect that this will work for the high energy tails of distributions.\footnote{It was recently shown in \cite{Grober:2017uho} that one could approximately reconstruct the non-analytic $m_t$-dependence from the threshold behavior of the amplitude using a Pad\'e ansatz, however with increasing uncertainties at higher energies.} If the energy is high enough, on the other hand, one could exploit the opposite limit $|s|,|t|,|u| \gg m_t^2$ to simplify the calculation. Here $s$, $t$, $u$ are the usual Mandelstam variables. In this limit, one can perform a double expansion in terms of $m_t^2/s$ and $m_h^2/m_t^2$, where $m_h$ is the mass of the Higgs boson. This has been done for the two-loop amplitudes in $H+j$ production in \cite{Kudashkin:2017skd, Lindert:2018iug}, and for the two-loop planar integrals in Higgs boson pair production in \cite{Davies:2018ood}. It should be noted that the $m_t \to 0$ limit is not regular, and the expansion in $m_t^2/s$ is not a Taylor series. Instead, one generally encounters powers of $\ln(m_t^2/s)$ in the expansion. This method is efficient for the high energy tails of differential distributions, e.g., in the region of very large Higgs transverse momentum or the region of very large Higgs pair invariant mass.

Recently, a new method was proposed in \cite{Bonciani:2018omm} to calculate the NLO QCD corrections to Higgs boson pair production, in which the integrals are expanded in the limit of small transverse momentum (compared to $\sqrt{s}$ and $m_t$). The resulting integrals are functions of $s$ and $m_t^2$, and can be calculated analytically. This method provides a rather good approximation for the bulk of the cross section $\sqrt{s} \lesssim \unit{750}{\GeV}$. However, it cannot be applied to the high energy tails of differential distributions, as the expansion becomes divergent in the region $\sqrt{s} \gtrsim \unit{900}{\GeV}$ or in the region of large transverse momentum.

Given the above expansion methods valid in several distinct kinematic regions, it is interesting to ask whether it is possible to construct an expansion which is valid in the whole phase space. The answer is yes, as will be outlined in this article, with the price that the resulting integrals are more complicated (but are still simpler than the original integrals, and are possible to be calculated with the help of differential equations). Our starting point is an expansion in the limit of small $m_h$, without assumptions on the other scales $s$, $t$, $u$ and $m_t$. The resulting integrals involve only massless external legs and resemble those appearing in, e.g., the di-jet production process \cite{Becchetti:2017abb}. For these integrals, we first perform the standard integration by parts (IBP) reduction into a set of master integrals \cite{Chetyrkin:1981qh}, and then construct differential equations for the master integrals \cite{Kotikov:1990kg, Kotikov:1991pm}. Wherever possible, we convert the differential equations into a canonical form \cite{Henn:2013pwa} via an appropriate basis choice, and obtain the solutions in terms of the Chen iterated integrals \cite{Chen:1977oja}. This is the case for 3 out of 4 integral families in Higgs boson pair production. As for the remaining integral family, we reduce the system of differential equations such that at most two master integrals are coupled at the leading order in the dimensional regulator. This allows a solution in terms of elliptic integrals.

The method developed along this line can be viewed as a unification of the existing expansion-based approaches, and is valid in a broader region of phase space. As such, it represents an improvement over the other methods, especially in the phenomenologically important intermediate regions. Our method is not restricted to Higgs boson pair production and $H+j$ production.\footnote{Note that there have been progresses to analytically calculate the planar two-loop integrals for $H+j$ production using the method of differential equations \cite{Bonciani:2016qxi}.} It can be applied to any process involving a heavy quark loop, such as the top quark loop contribution to $gg \rightarrow HZ$ \cite{Hasselhuhn:2016rqt} and $gg \rightarrow ZZ$ \cite{Campbell:2016ivq}, as well as the mixed QCD-electroweak corrections to $e^+ e^- \to HZ$ \cite{Gong:2016jys, Sun:2016bel}.

The article is organized as follows. In Section~\ref{sec:expansion}, we introduce our method to derive the small Higgs mass expansion, and demonstrate its validity using the known one-loop results. In Section~\ref{sec:twoloop}, we apply our method to the two-loop non-planar integrals in Higgs boson pair production, and show the comparison of our results in one of the topologies against the numerical results from sector decomposition. We conclude and discuss future developments of our method in Section~\ref{sec:conclusion}. And finally in the Appendix, we list the basis of master integrals we use in our calculation.

\section{Expansion in terms of external Higgs masses}
\label{sec:expansion}

In this and the following sections, we will use Higgs boson pair production as the concrete example to demonstrate our method, namely, we consider two-loop contributions to the process
\begin{align}
g(p_1) + g(p_2) \to H(p_3) + H(p_4) \, ,
\end{align}
where the kinematic invariants are
\begin{align}
\label{eq:kin_massive}
s = (p_1+p_2)^2 \, , \quad t = (p_1-p_3)^2 \, , \quad u = (p_2-p_3)^2 \, , \quad p_1^2=p_2^2=0 \, , \quad p_3^2=p_4^2=m_h^2 \, .
\end{align}
They satisfy the usual relation $s+t+u=2m_h^2$. As a result, the scattering amplitude depends on 4 energy scales which can be chosen as $s$, $t$, $m_h$ and $m_t$, where the top quark mass enters through propagators. 

The presence of multiple scales in the two-loop amplitude makes it rather difficult to calculate. On one hand, it is highly non-trivial to reduce the amplitude into a set of master integrals via the usual IBP method. On the other hand, many of the master integrals are not expected to have a representation in terms of (generalized) polylogarithms or even elliptic integrals. Given this situation, the various approximation methods mentioned in the introduction exploit different kinematic limits to reduce the number of scales in the problem. This simplifies both the reduction of the amplitude and the evaluation of the master integrals. For example, the large $m_t$ expansion corresponds to the limit $m_t^2 \gg |s|,|t|,m_h^2$; the large energy expansion corresponds to $|s|,|t| \gg m_t^2 \gg m_h^2$; and the small $p_T$ expansion corresponds to $|s|,m_t^2 \gg |t|,m_h^2$. Note that all the above expansions can be obtained by first expanding around the limit $m_h^2 \to 0$, and then further dealing with the 3 remaining scales $s$, $t$ and $m_t^2$. This small Higgs mass limit is therefore more general and is valid in a broader region of phase space.

In general, loop integrals may develop new singularities in the limit where one of the internal or external masses is taken to zero. If that's the case, the expansion around that limit will not be a normal power series. An example is the limit of small top quark mass discussed in \cite{Davies:2018ood}, where the expansion involves powers of $\log(m_t)$ in addition to powers of $m_t$. However, external Higgs bosons are special, since they only couple to \emph{massive} particles directly. As a result, no new singularities arise in the limit $m_h \to 0$, and we can Taylor-expand a generic integral as\footnote{Subtleties arise when more than two massless external partons are present, e.g., in $H + j$ production \cite{Kudashkin:2017skd}. In such cases certain integrals are singular in the limit $m_h \to 0$. However, the full amplitude remains finite in that limit, and the expansion of the amplitude is well-behaved.} 
\begin{align}
\label{eq:expansion}
I(s,t,m_t^2,m_h^2,\epsilon) = \sum_{n=0}^{\infty} \frac{m_h^{2n}}{n!} \, I^{(n)}(s,t,m_t^2,\epsilon) \, ,
\end{align}
where
\begin{align}
I^{(n)}(s,t,m_t^2,\epsilon) = \partial_{m_h^2}^n I(s,t,m_t^2,m_h^2,\epsilon) \bigg|_{m_h^2=0} \, .
\end{align}
The above expansion coefficients can be obtained by calculating integrals with only massless external legs, which are simpler than the original ones.

To demonstrate our method, we take the one-loop integrals appearing in Figure~\ref{fig:LO}(a) as an example. The propagators are given by
\begin{align}
D_1 = k^2-m_t^2, \; D_2 =(k+p_1)^2-m_t^2, \; D_3 = (k+p_1+p_2)^2-m_t^2, \; D_4 = (k+p_3)^2-m_t^2,
\end{align}
where $k$ is the loop momentum. We define the family of integrals
\begin{align}
I_{a_1,a_2,a_3,a_4}(s,t,m_t^2,m_h^2,\epsilon) \equiv \frac{16\pi^2}{i} \bigg( \frac{m_t^2}{4\pi} \bigg)^\epsilon \, \Gamma(1+\epsilon)  \int \frac{d^d k}{(2\pi)^d} \frac{1}{D_1^{a_1} \, D_2^{a_2} \, D_3^{a_3} \, D_4^{a_4}} \, ,
\end{align}
where $d=4-2\epsilon$ is the space-time dimension in dimensional regularization. In order to perform the expansion, we need to take the derivative of the above integrals with respect to $m_h^2$. This can be accomplished via the following operator
\begin{align}
\label{eq:derivative}
\partial_{m_h^2} = \frac{-m_h^2 \, p_1^\mu + t \, p_2^\mu + (m_h^2-t) \, p_3^\mu}{m_h^4 - 2 m_h^2 t + t (s+t)} \, \partial_{p_3^\mu} \, .
\end{align}
As an example, we have
\begin{align}
\partial_{m_h^2} I_{1,1,1,1} &= \frac{1}{m_h^4 - 2 m_h^2 t + t (s+t)} \Big[ m_h^2 I_{1,0,1,2} - m_h^2 I_{1,1,1,1} - t^2 I_{1,1,1,2} \nonumber
\\
&\hspace{4em} -t \left( -m_h^2 I_{1,1,1,2} + I_{0,1,1,2} - I_{1,0,1,2} + I_{1,1,0,2} - I_{1,1,1,1} \right) \Big] \, ,
\end{align}
where we have suppressed the arguments of the integrals. The expansion of $I_{1,1,1,1}$ can then be written as
\begin{align}
\label{eq:exp1l1}
I_{1,1,1,1} = \tilde{I}_{1,1,1,1} + \frac{m_h^2}{s+t} \left[ -t \tilde{I}_{1,1,1,2} -  \left( \tilde{I}_{0,1,1,2} - \tilde{I}_{1,0,1,2} + \tilde{I}_{1,1,0,2} - \tilde{I}_{1,1,1,1} \right) \right] + \mathcal{O}(m_h^4) \, ,
\end{align}
where
\begin{align}
\tilde{I}_{a_1,a_2,a_3,a_4}(s,t,m_t^2,\epsilon) = \lim_{m_h^2 \to 0} I_{a_1,a_2,a_3,a_4}(s,t,m_t^2,m_h^2,\epsilon) \, .
\end{align}
Note that no reduction has been performed at this stage, which is important since the IBP reduction for the original $m_h$-dependent integrals can become very complicated at the two-loop order. The reduction can be carried out for the $\tilde{I}$ integrals when necessary, which is much easier to do. For example, after reduction, Eq.~(\ref{eq:exp1l1}) can be simplified to
\begin{align}
I_{1,1,1,1} = \tilde{I}_{1,1,1,1} + \frac{m_h^2}{s+t} \left[ -t \tilde{I}_{1,1,1,2} -  \left( 2 \tilde{I}_{1,1,0,2} - \tilde{I}_{1,0,1,2} - \tilde{I}_{1,1,1,1} \right) \right] + \mathcal{O}(m_h^4) \, .
\end{align}
Note that we have chosen the basis of master integrals such that the above formula is simple to show. It is of course straightforward to convert to the conventional basis of one-loop scalar integrals, which we employ for our numerical computations.
We have only shown the expansion up to order $m_h^2$. There is no difficulty in extending the expansion to higher powers of $m_h$. In practice we find that keeping terms up to $m_h^4$ or $m_h^6$ already provides rather good approximations to the exact results. This will be clear from the numerical results shown in the following.

\begin{figure}[t!]
\centering
\includegraphics[width=0.48\textwidth]{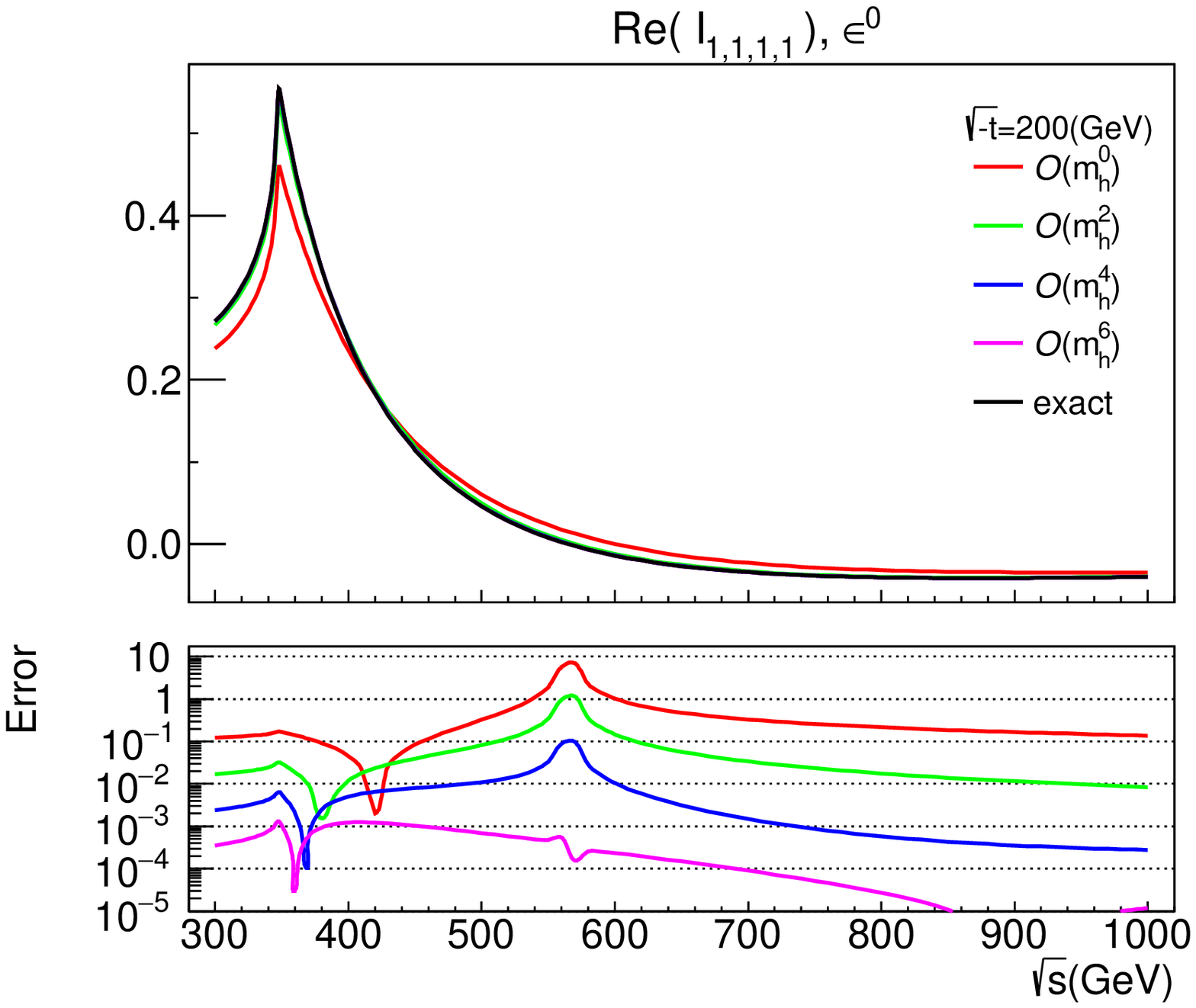}
\includegraphics[width=0.48\textwidth]{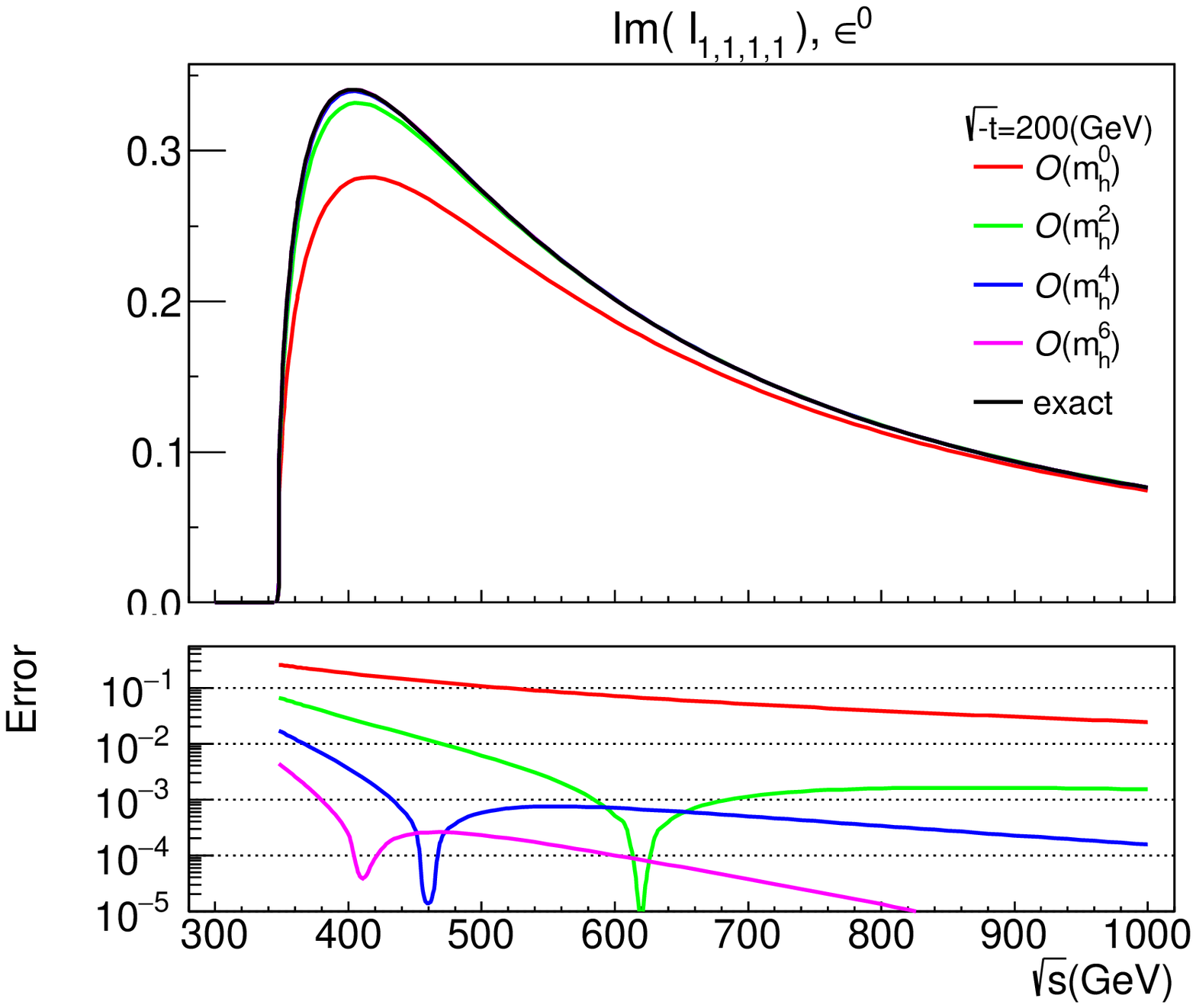}
\\[-10ex]
\includegraphics[width=0.48\textwidth]{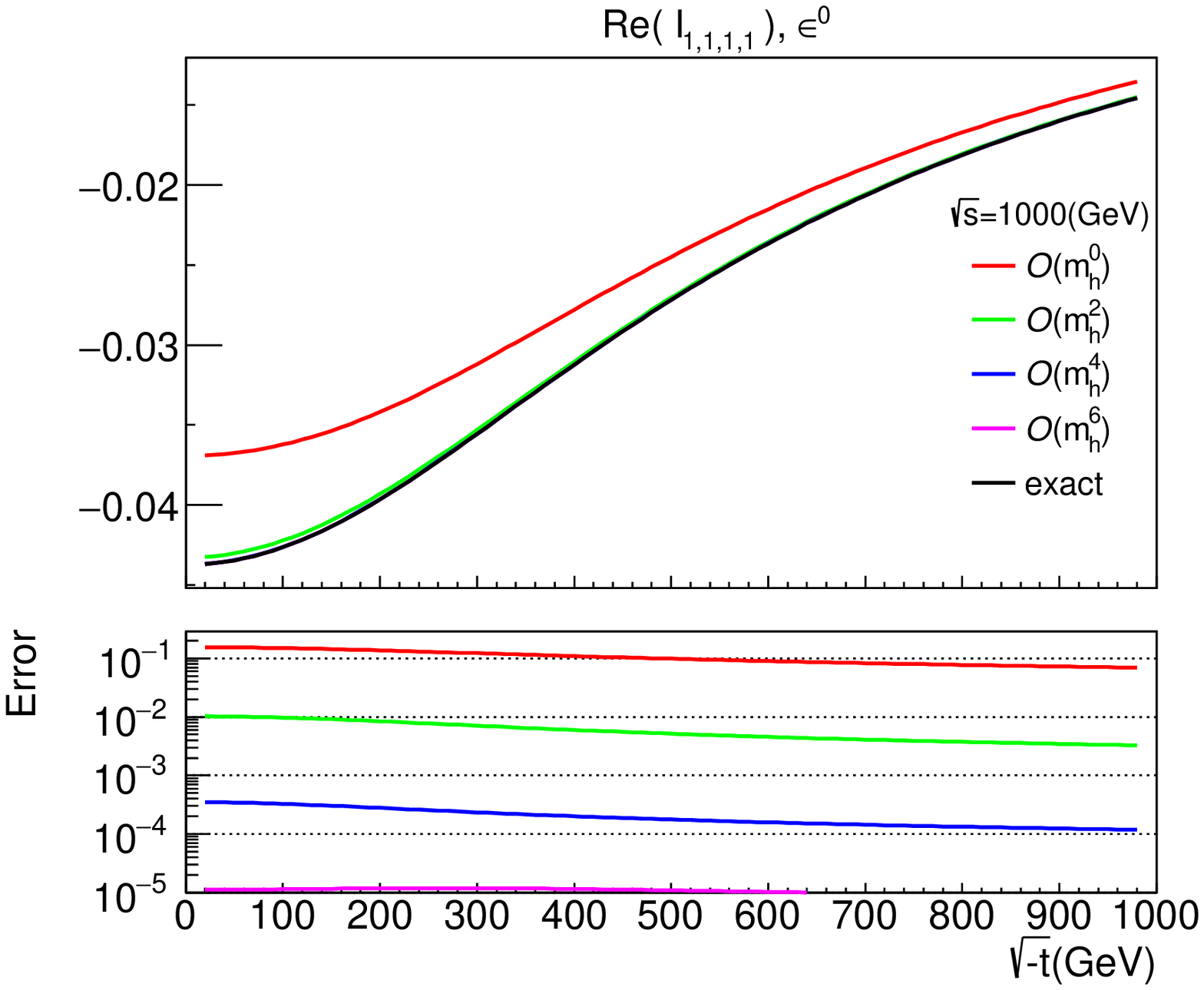}
\includegraphics[width=0.48\textwidth]{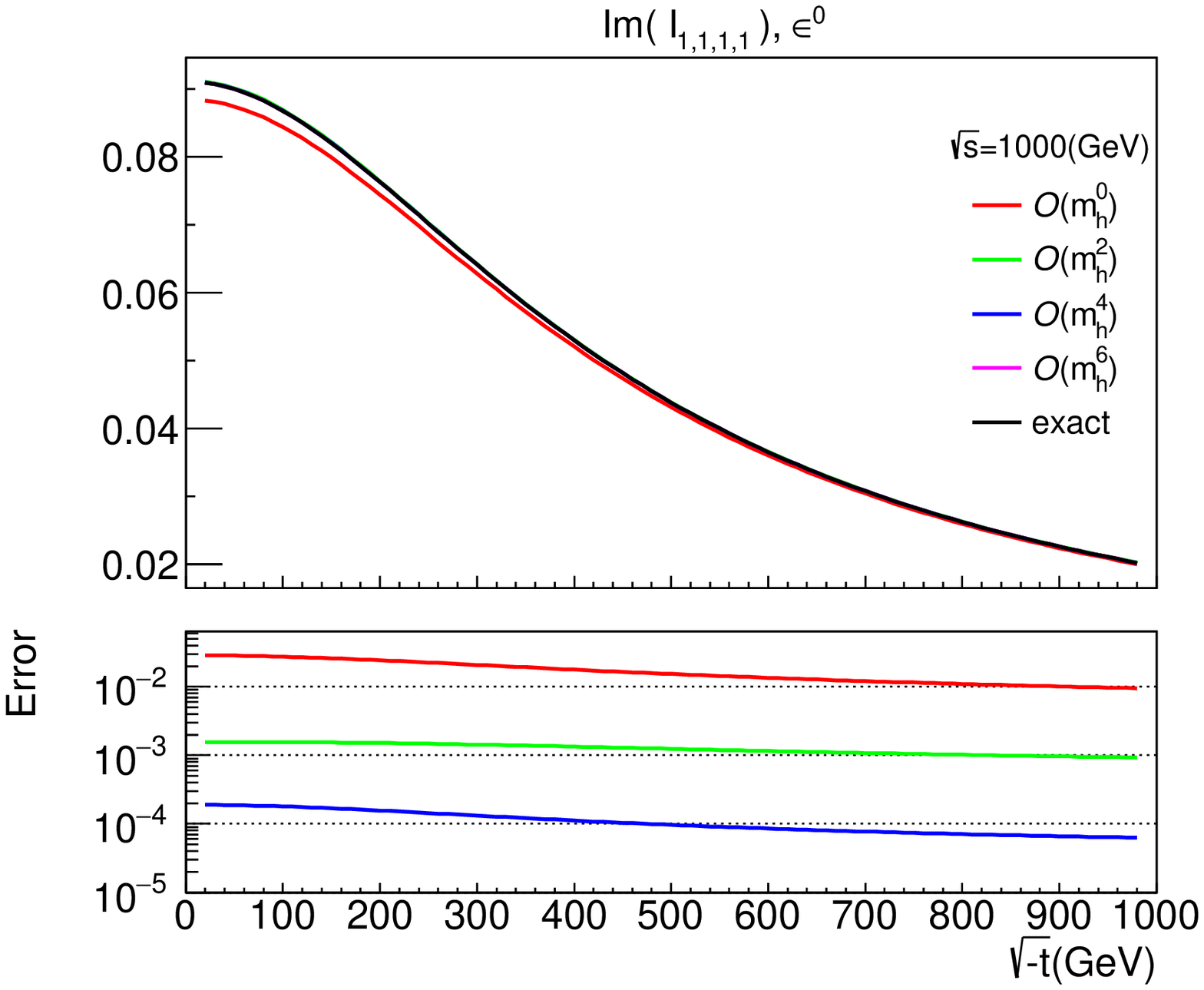}
\vspace{-10ex}
\caption{\label{fig:I1111}The real part (left plots) and the imaginary part (right plots) of the order $\epsilon^0$ coefficient of the one-loop integral $I_{1,1,1,1}$. The upper plots fix $\sqrt{-t}=\unit{200}{\GeV}$ and show the integral as a function of $\sqrt{s}$, while the lower plots fix $\sqrt{s}=\unit{1000}{\GeV}$ and show the integral as a function of $\sqrt{-t}$.
In the lower panels of each plot, we show the relative errors of the approximate results against the exact result (see also the text for definition).
The integral has been multiplied by $m_t^4$ to make it dimensionless.}
\end{figure}

We choose the masses to be $m_t=\unit{173.3}{\GeV}$ and $m_h=\unit{125.1}{\GeV}$, and vary the kinematic variables $s$ and $t$ to see the goodness of the approximation in different regions of phase space. We first show the result for the integral $I_{1,1,1,1}$. For convenience we rescale the integral by an appropriate power of $m_t$ such that the result is a dimensionless number. In the upper plots of Fig.~\ref{fig:I1111}, we fix $t=-(\unit{200}{\GeV})^2$ and show the real part and the imaginary part of the order $\epsilon^0$ coefficient as a function of $\sqrt{s}$.
We also show the relative errors of the approximate results against the exact result, defined as
\begin{equation}
\text{Error} = \left| \frac{\text{Approximate} - \text{Exact}}{\text{Exact}} \right| .
\end{equation}
It can be seen that the expansion up to order $m_h^4$ already gives sub-percent accuracies for both the real and the imaginary parts for almost all values of $\sqrt{s}$, ranging from the threshold region $\sqrt{s} \gtrsim 2m_h$, to the $t\bar{t}$ threshold $\sqrt{s} \sim 2m_t$, to the high energy regime $\sqrt{s} \gg 2m_t$. The only exception is the real part at around $\sqrt{s} \sim \unit{560}{\GeV}$, where it happens by coincidence that the value of the integral is close to zero. In such cases one needs to add the order $m_h^6$ term, which leads to per-mille accuracy in all regions of phase space. Similar behavior can be observed in the lower plots of Fig.~\ref{fig:I1111}, where we fix $\sqrt{s}=\unit{1000}{\GeV}$ and show the integral as a function of $\sqrt{-t}$. Here the approximation at order $m_h^4$ gives better-than-per-mille accuracy in the whole range and it is not necessary to include the $\mathcal{O}(m_h^6)$ corrections.

The behavior of a single integral is perhaps not convincing enough. We now turn to investigate the partonic (differential) cross sections which are physically more relevant. We start by writing the amplitude as 
\begin{align}
\mathcal{M}^{\mu\nu}_{ab} = \frac{G_F}{\sqrt{2}} \frac{\alpha_s}{2\pi} \, s \, \delta_{ab} \, \Big[ A_1^{\mu\nu} F_1(s,t,m_t^2,m_h^2) + A_2^{\mu\nu} F_2(s,t,m_t^2,m_h^2) \Big] \, ,
\end{align}
where the two tensor structures are given by \cite{Plehn:1996wb}
\begin{align}
A_1^{\mu\nu} &= g^{\mu\nu} - \frac{p_1^\nu \, p_2^\mu}{p_1 \cdot p_2} \, , \nonumber
\\
A_2^{\mu\nu} &= g^{\mu\nu} + \frac{m_h^2 \, p_1^\nu \, p_2^\mu}{p_T^2 \; p_1 \cdot p_2} - \frac{2 \, p_2 \cdot p_3 \; p_1^\nu \, p_3^\mu + 2 \, p_1 \cdot p_3 \; p_2^\mu \, p_3^\nu}{p_T^2 \; p_1 \cdot p_2} + \frac{2 \, p_3^\mu \, p_3^\nu}{p_T^2} \, ,
\end{align}
where $p_T$ denotes the transverse momentum of the top quark and can be written as
\begin{align}
p_T^2 = \frac{2 \, p_1 \cdot p_3 \; p_2 \cdot p_3}{p_1 \cdot p_2} - m_h^2 \, .
\end{align}
At one-loop, the two form factors $F_1$ and $F_2$ can be evaluated either exactly or using the small $m_h$ expansion. They can then be used to calculate the partonic differential cross section
\begin{align}
\frac{d\hat{\sigma}}{dp_T} = \frac{G_F^2\alpha_s^2 p_T\sqrt{s}}{1024\pi^3\sqrt{s-4(p_T^2+m_h^2)}} \Big( |F_1|^2 + |F_2|^2 \Big) \, ,
\end{align}
and also the partonic total cross section $\hat{\sigma}$ by integrating over $p_T$.

\begin{figure}[t!]
\centering
\includegraphics[width=0.48\textwidth]{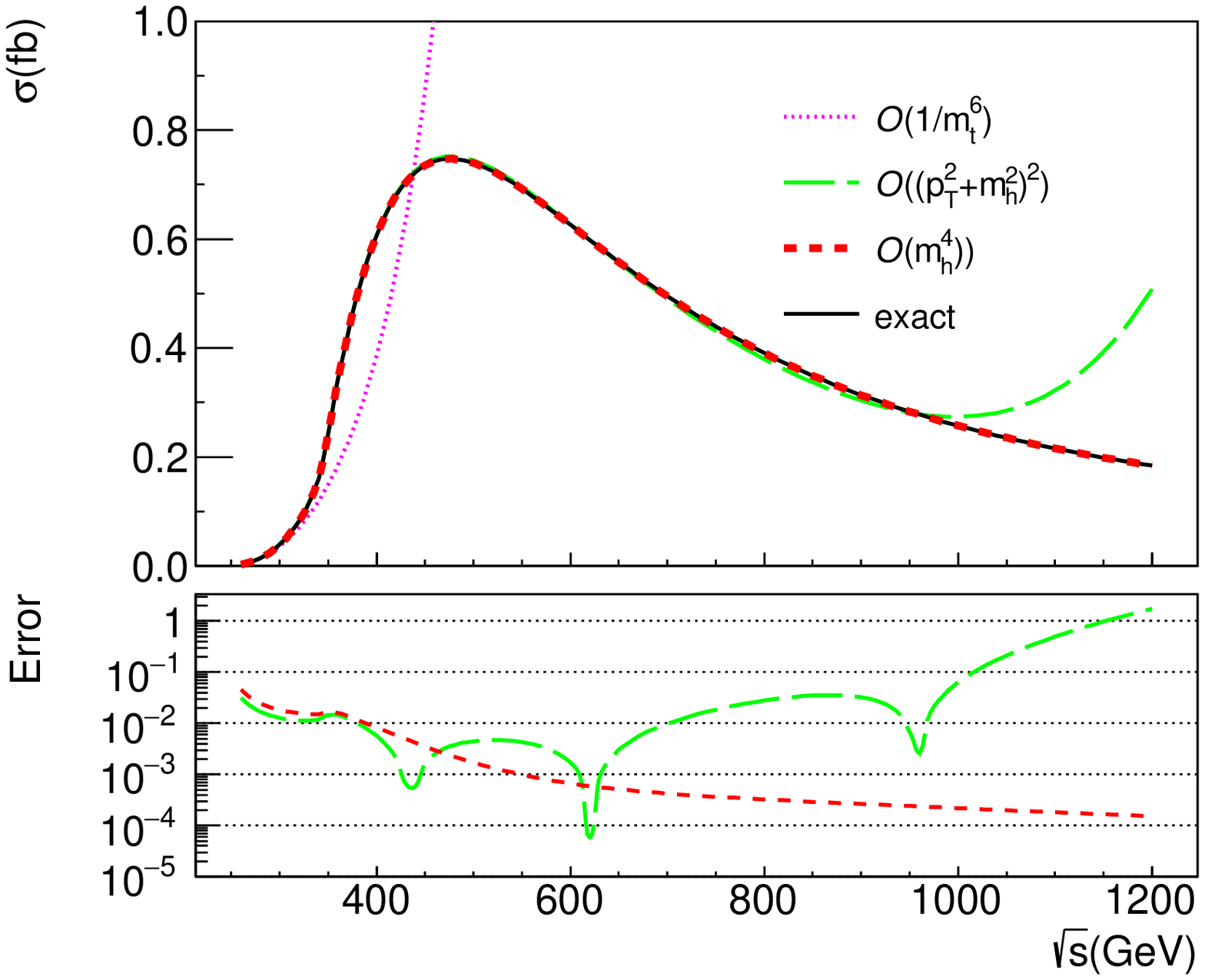}
\includegraphics[width=0.48\textwidth]{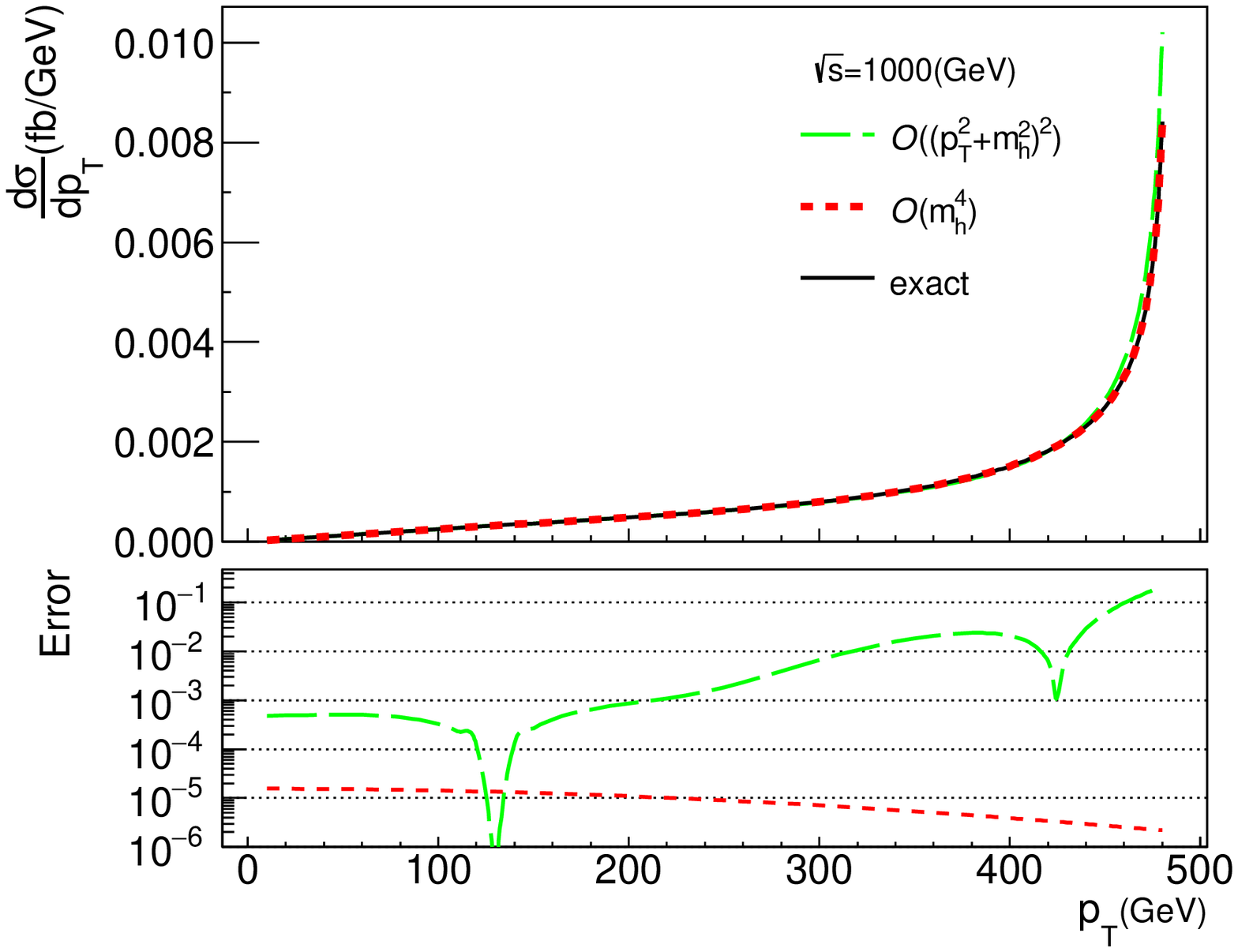}
\vspace{-12ex}
\caption{\label{fig:LOcs}Left side: the partonic total cross section as a function of $\sqrt{s}$. Right side: the transverse momentum distribution of the Higgs boson at the parton level with $\sqrt{s}=\unit{1000}{\GeV}$.}
\end{figure}

At this point, it is interesting to compare our approximation to the other methods, e.g., the $1/m_t$ expansion in \cite{Grigo:2013rya} and the $p_T^2/s$ expansion in \cite{Bonciani:2018omm}. On the left side of Fig.~\ref{fig:LOcs}, we show the partonic total cross section $\hat{\sigma}$ as a function of $\sqrt{s}$. The black solid line is the exact result, while the other curves represent three different approximations. It is clear that the large-$m_t$ expansion only works in the region $s < 4m_t^2$, as expected. The $p_T^2/s$ expansion is valid in a broader range, and provides a reasonable approximation to the exact result up to $\sqrt{s} \lesssim \unit{900}{\GeV}$. However, going beyond that, the $p_T^2/s$ expansion quickly becomes divergent. On the other hand, our small-$m_h$ expansion works perfectly across the whole range. To see more clearly the behaviors of the small-$p_T$ expansion and our small-$m_h$ expansion, in the lower panel of the plot we show the relative error with respect to the exact result. We find that the qualities of the two approximations are similar for $\sqrt{s} < \unit{500}{\GeV}$. Beyond that, the small-$p_T$ expansion becomes worse and worse, while the small $m_h$ expansion becomes better and better, and provides a better-than-per-mille approximation to the exact result.

To see more clearly the difference between the $p_T^2/s$ expansion and the small-$m_h$ expansion at high energy, we show on the right side of Fig.~\ref{fig:LOcs} the transverse momentum distribution at the parton level with $\sqrt{s}=\unit{1000}{\GeV}$. We find that the accuracy of the small-$m_h$ expansion is at the level of $10^{-5}$ in the whole range of $p_T$. We also observe that the distribution peaks towards the right end, which means that the dominant contribution to the partonic total cross section comes from the high $p_T$ region. It is clear that the small-$p_T$ expansion cannot be a good approximation in this region, which is due to the fact that the condition $p_T \ll m_t$ is no longer fulfilled. This also explains why the small-$p_T$ expansion fails for the partonic total cross section at large $\sqrt{s}$, as observed from the left plot. 

The above discussions demonstrate the validity of the small-$m_h$ expansion in the entire phase space at the one-loop level. This makes us confident that the same will be true at higher loop orders. In the following section, we apply our expansion to the two-loop amplitude, with the goal to provide a fast and reliable method to evaluate the NLO QCD corrections to Higgs boson pair production.

\section{Expansion at the two-loop order}
\label{sec:twoloop}

\subsection{Setup}

\begin{figure}[t!]
\centering
\includegraphics[width=11cm,height=4cm]{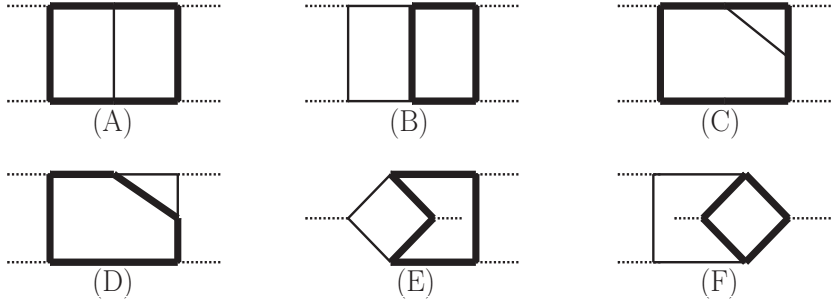}
\vspace{-1ex}
\caption{Topologies relevant to the NLO QCD corrections to Higgs boson pair production after expansion in the small $m_h$ limit. The thick lines represent massive propagators (top quarks), while the thin lines represent massless propagators (gluons). The external legs (dashed lines) are all light-like.}
\label{fig:topologies}
\end{figure}

We now turn to the NLO (two-loop) QCD corrections to Higgs boson pair production. The expansion in terms of $m_h^2$ takes the form as Eq.~(\ref{eq:expansion}) and can be performed using the derivative operator Eq.~(\ref{eq:derivative}). We stress that this can be done at the amplitude level, without the need of reduction beforehand. We have carried out the expansion up to order $m_h^4$. The extension to higher powers in $m_h$ is straightforward. The expansion coefficients can be obtained by calculating integrals with massless external legs. After applying crossing symmetries, all the integrals can be classified into 6 integral families. They corresponds to the 6 topologies depicted in Fig.~\ref{fig:topologies}.

We employ the IBP identities to reduce the integrals in these topologies into master integrals. It happens that after reduction, all the 7-propagator integrals in topology C and D can be expressed in terms of integrals in sub-topologies with 6 propagators of less. All these sub-topologies also appear in topology A and B, so that we don't need to calculate them again.
We therefore only need to consider 4 integral families. We first define the $m_h$-dependent integrals
\begin{align}
I_{\{a_i\}}(s,t,m_t^2,m_h^2,\epsilon) = \Bigg[ \frac{16\pi^2}{i} \bigg( \frac{m_t^2}{4\pi} \bigg)^\epsilon \, \Gamma(1+\epsilon) \Bigg]^{2} \int \frac{d^d k_1}{(2\pi)^d} \frac{d^d k_2}{(2\pi)^d} \prod_{i=1}^9 \frac{1}{D_i^{a_i}} \, ,
\label{eq:twoloopint}
\end{align}
where $k_1$ and $k_2$ are loop momenta, and $\{a_i\}$ denotes the collection of powers $a_i$ on the propagators $D_i$. We then define
\begin{align}
\tilde{I}_{\{a_i\}}(s,t,m_t^2,\epsilon) = \lim_{m_h^2 \to 0} I_{\{a_i\}}(s,t,m_t^2,m_h^2,\epsilon) \, ,
\end{align}
which are the main objects to be calculated in this section. The 4 relevant integral families are defined by their corresponding propagators as the following:
\begin{align}
\text{A} &: \quad \big\{ k_1^2-m_t^2, (k_1+p_1)^2-m_t^2, (k_1+p_1+p_2)^2-m_t^2, (k_1+k_2)^2, k_2^2-m_t^2, \nonumber
\\
&\qquad (k_2-p_3)^2-m_t^2, (k_2-p_1-p_2)^2-m_t^2, (k_2-p_1)^2-m_t^2, (k_1+p_3)^2-m_t^2 \big\} \, , \nonumber
\\
\text{B} &: \quad \big\{ k_1^2, (k_1+p_1)^2, (k_1+p_1+p_2)^2, (k_1+k_2)^2-m_t^2, k_2^2-m_t^2, \nonumber
\\
&\qquad (k_2-p_3)^2-m_t^2, (k_2-p_1-p_2)^2-m_t^2, (k_2-p_1)^2-m_t^2, (k_1+p_3)^2 \big\} \, , \nonumber
\\
\text{E} &: \quad \big\{ k_1^2, (k_1+p_1)^2, (k_1+k_2)^2-m_t^2, k_2^2-m_t^2, (k_2-p_3)^2-m_t^2, \nonumber
\\
&\qquad (k_2-p_1-p_2)^2-m_t^2, (k_1+k_2-p_2)^2-m_t^2, (k_2-p_1)^2-m_t^2, (k_1-p_3)^2 \big\} \, ,\nonumber
\\
\text{F} &: \quad \big\{ (k_1-p_1)^2, k_1^2, (k_1+p_2)^2, (k_1+k_2-p_1)^2-m_t^2, k_2^2-m_t^2,  \nonumber
\\
&\qquad (k_2-p_3)^2-m_t^2, (k_1+k_2+p_2-p_3)^2-m_t^2, (k_1-p_3)^2, (k_2-p_1)^2-m_t^2 \big\} \, .
\label{eq:propagators}
\end{align}
Integrals in each of these families can be reduced to a set of master integrals. For that purpose we employ the program packages \texttt{FIRE5} \cite{Smirnov:2014hma} and \texttt{LiteRed} \cite{Lee:2012cn}. We find 29 master integrals in topology A, 32 for topology B, 54 for topology E, and 37 for topology F.

The kinematic invariants are defined as in Eq.~(\ref{eq:kin_massive}), with the exception that we now have $p_3^2=p_4^2=0$ and $s+t+u=0$. We choose $s$, $t$ and $m_t^2$ as independent scales and introduce the following dimensionless quantities
\begin{align}
\mu \equiv -\frac{4m_t^2}{s} \, , \quad \nu \equiv -\frac{4m_t^2}{t} \, .
\end{align}
Physically, we have $s > 2m_h^2$ and $t,u < 0$. When $s$ is above the $t\bar{t}$ threshold, namely $s > 4m_t^2$, some of the integrals will develop imaginary parts. For convenience, we will first work in the unphysical region
\begin{align}
s < 0 \, , \quad t < 0 \, , \quad -4m_t^2 < s + t < 0 \, ,
\end{align}
which corresponds to
\begin{align}
\mu > 1 \, , \quad \nu > 1 \, , \quad \mu + \nu > 4 \, .
\end{align}
This guarantees that all the integrals are real. After obtaining the expressions of the master integrals, we can perform an analytic continuation to the physical region and then evaluate them numerically.

Among the 4 integral families, the two planar topologies A and B have already been discussed in \cite{Caron-Huot:2014lda, Becchetti:2017abb}. In the following, we discuss the calculation of the master integrals in the two non-planar topologies E and F.

\subsection{Calculation of the master integrals for topology E}
\label{sec:topologyE}

\subsubsection{Analytic structures}

\begin{figure}[t!]
\centering
\includegraphics[width=13cm,height=10cm]{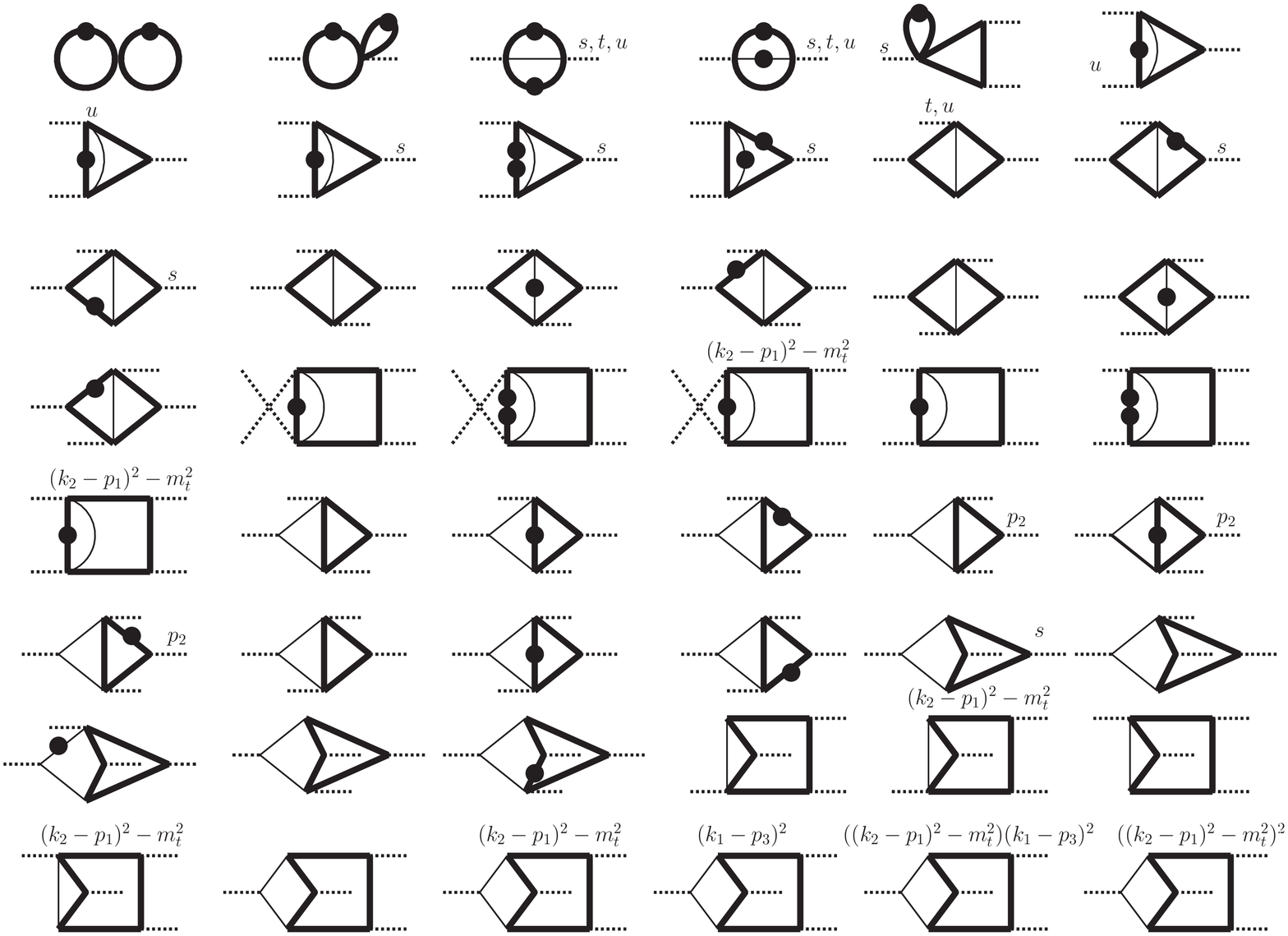}
\caption{Pre-canonical master integrals in topology E. The thick lines represent massive propagators (from top quarks) and the thin lines denote massless propagators (from gluons). The labels $s$, $t$ and $u$ on the external lines represent the (squared) momenta flowing through those legs. The external lines without labels have light-like momenta.}
\label{fig:pre-canonical}
\end{figure}

Topology E is the simpler one in the two non-planar topologies, in that it is possible to cast the differential equations satisfied by the master integrals into a canonical form \cite{Henn:2013pwa}. For that purpose we use a method similar to the one used in \cite{Argeri:2014qva}. We start from the sub-topologies with the lowest number of propagators. We choose appropriate pre-canonical master integrals so that the differential equations are simple enough and are of the form
\begin{align}
\frac{\partial}{\partial x_i} \vec{f}_0(\vec{x},\epsilon) = \left[ \epsilon \tilde{A}_{0i}(\vec{x}) + \tilde{B}_{0i}(\vec{x}) \right] \vec{f}_0(\vec{x},\epsilon) \, ,
\label{differential equation}
\end{align}
where the vector $\vec{f}_0(\vec{x},\epsilon)$ denotes the collection of the pre-canonical master integrals, and $\vec{x}$ is the collection of independent kinematic variables (in our case $\vec{x} = \{\mu,\nu\}$. $\tilde{A}_{0i}(\vec{x})$ and $\tilde{B}_{0i}(\vec{x})$ are two square matrices which do not depend on the dimensional regulator $\epsilon$. We then apply a linear transform $T(\vec{x},\epsilon$) on the vector $\vec{f}_0(x,\epsilon)$ such that the new vector $\vec{f}(\vec{x},\epsilon) = T(\vec{x},\epsilon) \vec{f}_0(\vec{x},\epsilon)$ satisfies a system of differential equations in the canonical form
\begin{align}
\label{eq:canonical}
\frac{\partial}{\partial x_i} \vec{f}(\vec{x},\epsilon) = \epsilon \tilde{A}_i(\vec{x}) \vec{f}(\vec{x},\epsilon) \, ,
\end{align}
where the matrices $\tilde{A}_i(\vec{x})$ are at most algebraic functions of the variables $x_i$. We then proceed to topologies with the number of propagators higher by 1, and repeat the above process. Finally we arrive at the top-level topology with 7 propagators. In Fig.~\ref{fig:pre-canonical} we list the diagrammatic representations of the pre-canonical integrals in topology E. The transformation of them into the canonical basis is discussed in the Appendix. 

Given the differential equation in the canonical form (\ref{eq:canonical}), the solution for the master integrals can be generically written as Chen iterated integrals \cite{Chen:1977oja}. For that purpose, it is convenient to rewrite the differential equations as
\begin{align}
d\vec{f}(\vec{x},\epsilon) = \epsilon \, dA(\vec{x}) \, \vec{f}(\vec{x},\epsilon) \, ,
\end{align}
where $dA(\vec{x})=\sum_i \tilde{A}_i(\vec{x}) \, dx_i$ which can be expressed in the d-log form
\begin{align}
dA(\vec{x}) = \sum_k A_k \, d\log\alpha_k(\vec{x}) \, .
\end{align}
Here for a given $k$, $A_k$ is a constant matrix independent of the kinematic variables, while $\alpha_k(\vec{x})$ is an algebraic function of the kinematic variables and is called a ``letter''. The collection of all letters is called the ``alphabet'', which completely determines the class of functions appearing in the master integrals $\vec{f}(\vec{x},\epsilon)$. The general solution can be written as
\begin{align}
\label{eq:pathint}
\vec{f}(\vec{x},\epsilon) = \mathcal{P} \exp \bigg[ \epsilon \int_{\vec{x}_0}^{\vec{x}} dA(\vec{x}') \bigg] \, \vec{f}(\vec{x}_0,\epsilon)
\end{align}
where $\mathcal{P}$ denotes path-ordering along the path connecting the boundary point $\vec{x}_0$ and the destination $\vec{x}$. The boundary conditions $\vec{f}(\vec{x}_0,\epsilon)$ can possibly be fixed by the analytic structure of the differential equations, or can be calculated directly.

The formal solution (\ref{eq:pathint}) is exact in $\epsilon$, while in practice one usually needs its expansion around the 4-dimensional limit $\epsilon = 0$. In order to do that, it is convenient to normalize the master integrals to have the property of uniform transcendental weights. The concept of transcendental weight (we will simply call it ``weight'' in the following) is closely related to iterated integrals. The weight of an algebraic number is defined to be 0, the weight of $\pi$ is defined to be 1, while the weight of the Riemann zeta value $\zeta_n$ is $n$. Given a weight-$n$ function $g(\vec{x})$, the weight of the integral
\begin{align}
\int_{\vec{x}_0}^{\vec{x}} g(\vec{x}') \, d\log(\alpha(\vec{x}'))
\end{align}
is defined to be $n+1$, where $\alpha(\vec{x})$ is an algebraic function of the kinematic variables. With this definition, it is clear that the $n$-fold iterated integral of the form
\begin{align}
F(\vec{x}) = \int_{\vec{x}_0}^{\vec{x}} d\log(\alpha_n(\vec{x}_n)) \cdots \int_{\vec{x}_0}^{\vec{x}_3} d\log(\alpha_2(\vec{x}_2))  \int_{\vec{x}_0}^{\vec{x}_2} d\log(\alpha_1(\vec{x}_1))
\label{iterated integral}
\end{align}
has transcendental weight $n$.

Now considering the expansion of the master integrals around $\epsilon=0$,
\begin{align}
\vec{f}(\vec{x},\epsilon) = \sum_{i=0}^\infty \vec{f}^{(i)}(\vec{x}) \, \epsilon^{i} \, .
\end{align}
We will normalize the master integrals such that the components of the vector $\vec{f}^{(i)}(\vec{x})$ are all weight-$i$ functions (or numbers). This is possible since they satisfy the canonical-form differential equation (\ref{eq:canonical}). For topology E, the prefactors for the normalization are collected in the Appendix. After normalization, the boundary conditions are simply given by
\begin{align}
\lim_{\mu,\nu\rightarrow \infty}f_i(\mu,\nu,\epsilon)=\delta_{i,1} \, ,
\end{align}
where the boundary $\mu,\nu \to \infty$ corresponds to $s,t \to 0$. The coefficient functions $\vec{f}^{(i)}(\vec{x})$ can then be written as iterated integrals order-by-order:
\begin{align}
\vec{f}^{(0)}(\vec{x}) &= \vec{f}^{(0)}(\vec{x}_0) \, , \nonumber
\\
\vec{f}^{(i)}(\vec{x}) &= \int_{\vec{x}_0}^{\vec{x}} dA(\vec{x}') \, \vec{f}^{(i-1)}(\vec{x}') + \vec{f}^{(i)}(\vec{x}_0) \, .
\label{eq:fi}
\end{align}

Given the above formal solutions, it is still non-trivial to convert them to explicit functions such as logarithms, polylogarithms and multiple polylogarithms (MPLs) \cite{Goncharov:1998kja}. For this purpose, we will use the concept of ``symbol'' \cite{Goncharov:1998kja, Brown:2009qja, Goncharov:2010jf, Duhr:2011zq}, which maps the iterated integrals to their integration kernels. Taking the function $F(\vec{x})$ in Eq. (\ref{iterated integral}) as an example, it's mapped to the symbol
\begin{align}
\mathcal{S}(F(\vec{x})) = \alpha_1(\vec{x}) \otimes \alpha_{2}(\vec{x}) \otimes \cdots \otimes \alpha_n(\vec{x}) \, .
\end{align}
The symbols of the iterated integrals in Eq.~(\ref{eq:fi}) can be written as
\begin{align}
\mathcal{S}(f_n^{(i)}(\vec{x})) = \sum_m \mathcal{S}(f_m^{(i-1)}(\vec{x}))
\otimes\mathcal{S}(A_{nm}(\vec{x}))\, .
\end{align}
The symbols satisfy a lot of algebraic relations which are of great help to simplify the complicated expressions. For example
\begin{align}
\alpha_{1}(\vec{x}) \otimes \cdots \otimes \big( \alpha_{i}(\vec{x}) \alpha_{i'}(\vec{x}) \big) \otimes \cdots \otimes \alpha_{n}(\vec{x}) &= \alpha_{1}(\vec{x}) \otimes \cdots \otimes \alpha_{i}(\vec{x}) \otimes \cdots \otimes \alpha_{n}(\vec{x}) \nonumber
\\
&+\alpha_{1}(\vec{x}) \otimes \cdots \otimes  \alpha_{i'}(\vec{x}) \otimes \cdots \otimes \alpha_{n}(\vec{x}) \, ,
\\
\alpha_{1}(\vec{x}) \otimes \cdots \otimes \big( c\alpha_{i}(\vec{x}) \big) \otimes \cdots \otimes \alpha_{n}(\vec{x}) &= \alpha_{1}(\vec{x}) \otimes \cdots \otimes \alpha_{i}(\vec{x})\otimes \cdots \otimes \alpha_{n}(\vec{x})\, , \nonumber
\end{align}
where $c$ is a constant. After simplification, it is possible to find an explicit functional representation for each symbol. In particular, when all the letters $\alpha_k(\vec{x})$ appearing in a given symbol are rational functions, it is straightforward to represent the function as polylogarithms or MPLs, which are well-studied and allow fast numerical evaluations. For example
\begin{align}
\mathcal{S}(\mathrm{Li}_k(z)) = -(1-z) \otimes \underbrace{z \otimes \cdots \otimes z}_{k-1 } \, .
\end{align}
However, letters in the alphabet for Higgs boson pair production contain square roots
\begin{align}
\left\{ \sqrt{1+\beta_i}, \sqrt{1+\beta_i+\beta_j}, \sqrt{16+8\beta_i +\beta_i^2+16\beta_j} \right\}, \quad i\neq j \in \{1,2,3\} \, ,
\label{eq:sqrt}
\end{align}
where $\beta_1=\mu$, $\beta_2=\nu$ and $\beta_3= -\mu\nu/(\mu+\nu)$. They make it challenging to convert the formal solutions to explicit functional forms. Fortunately, up to weight 2, only the first two kinds of square roots appear. In particular, there are only 4 kinds of symbols appearing at weight 2:
\begin{align}
&\frac{\sqrt{\beta_i+1}-1}{\sqrt{\beta_i+1}+1}\otimes\beta_i \, , \quad \frac{\sqrt{\beta_i+1}-1}{\sqrt{\beta_i+1}+1}\otimes(\beta_i+1) \, , \nonumber
\\
&\frac{\sqrt{\beta_i+1}-1}{\sqrt{\beta_i+1}+1}\otimes\frac{\sqrt{\beta_i+1}-\sqrt{\beta_i+\beta_j+1}}{\sqrt{\beta_i+1}+\sqrt{\beta_i+\beta_j+1}} \, , \quad h(\beta_i,\beta_j)\otimes h(\beta_i,\beta_j) \, ,
\end{align}
where $h(\beta_i,\beta_j)$ is some function of $\beta_i$ and $\beta_j$. The functional representation for the last symbol is simple:
\begin{align}
h(\beta_i,\beta_j) \otimes h(\beta_i,\beta_j) \rightarrow \frac{1}{2} \log^2(h(\beta_i,\beta_j)) \, ,
\label{eq:fij}
\end{align}
while for the first 3, we can get rid of the square roots with appropriate changes of variables. For the first two symbols, we use
\begin{align}
\beta_i=\frac{4z_i}{(1-z_i)^2} \, .
\end{align}
We work in the region $0 < z_i < 1$ such that the resulting functional representation is single-valued. This corresponds to $\beta_i > 0$.\footnote{Note that we cannot make $\beta_1$, $\beta_2$ and $\beta_3$ to be positive at the same time. This will become a subtlety for symbols involving all three $\beta_i$'s simultaneously, but does not affect our discussion here.} We can then express $z_i$ in terms of $\beta_i$ as
\begin{align}
z_i = \frac{\sqrt{1+\beta_i}-1}{\sqrt{1+\beta_i}+1} \, .
\end{align}
The expressions for $\beta_i < 0$ can be found by analytic continuation. For the third symbol, we parameterize
\begin{align}
\beta_i = \frac{ \left(1-x_{ij}^2\right) \left(1-y_{ij}^2\right) }{ (x_{ij}-y_{ij})^2 } \, , \quad \beta_j = \frac{4 x_{ij}y_{ij}}{(x_{ij}-y_{ij})^2} \, ,
\end{align}
where we take $0 < y_{ij} < x_{ij} < 1$ which corresponds to $\beta_i > 0$ and $\beta_j > 0$. The inverse relation is given by
\begin{align}
x_{ij} = \frac{\sqrt{1+\beta_j}+1}{\sqrt{1+\beta_i}+\sqrt{1+\beta_i+\beta_j}} \, ,
\quad y_{ij} = \frac{\sqrt{1+\beta_j}-1}{\sqrt{1+\beta_i}+\sqrt{1+\beta_i+\beta_j}} \, .
\end{align}
Now we can employ the algebraic properties of the symbols to further simplify the expressions. For example
\begin{align}
\frac{\sqrt{\beta_i+1}-1}{\sqrt{\beta_i+1}+1}\otimes\beta_i = z_i \otimes \frac{4z_i}{(1-z_i)^2} = z_i \otimes z_i - 2 \left[ z_i \otimes (1-z_i) \right] ,
\end{align}
and similarly for the remaining two symbols. These symbols are simple enough, such that their functional representations can be found via direct integration. The results are
\begin{align}
&\frac{\sqrt{\beta_i+1}-1}{\sqrt{\beta_i+1}+1} \otimes \beta_i \rightarrow 2 \mathrm{Li}_2(1-z_i) + \frac{1}{2} \log^2(z_i) \, , \nonumber
\\
&\frac{\sqrt{\beta_i+1}-1}{\sqrt{\beta_i+1}+1} \otimes (\beta_i+1) \rightarrow 2\mathrm{Li}_2(1-z_i) + 2\mathrm{Li}_2(-z_i) + 2\log(z_i) \log(z_i+1) + \frac{\pi^2}{6} \, , \nonumber
\\
&\frac{\sqrt{\beta_i+1}-1}{\sqrt{\beta_i+1}+1} \otimes \frac{\sqrt{\beta_i+1}-\sqrt{\beta_i+\beta_j+1}}{\sqrt{\beta_i+1}+\sqrt{\beta_i+\beta_j+1}} \rightarrow \mathrm{Li}_2(-x_{ij}) - \mathrm{Li}_2(x_{ij}) - \log(x_{ij}) \log\frac{1-y_{ij}}{1+y_{ij}} \nonumber
\\
&\hspace{17em} - \mathrm{Li}_2(-y_{ij}) + \mathrm{Li}_2(y_{ij}) + \log(y_{ij}) \log\frac{1-y_{ij}}{1+y_{ij}} \, .
\label{eq:weight_two}
\end{align}

We now turn to the weight-3 and weight-4 parts of the solution. These will involve the third square root in Eq.~(\ref{eq:sqrt}). Although it is still possible to find explicit functional forms from the symbols, it is often rather difficult \cite{Bonciani:2016qxi}. Therefore, we write them as one-fold integrals over the weight-2 functions
\begin{align}
\vec{f}^{(3)}(\vec{x}) &= \int_{\vec{x}_0}^{\vec{x}} dA(\vec{x}_1)\vec{f}^{(2)}(\vec{x}_1)+\vec{f}^{(3)}(\vec{x}_0) \, , \nonumber
\\
\vec{f}^{(4)}(\vec{x})&= \int_{\vec{x}_0}^{\vec{x}}dA(\vec{x}_2)\int_{\vec{x}_0}^{\vec{x}_2} dA(\vec{x}_1)\vec{f}^{(2)}(\vec{x}_1)+\int_{\vec{x}_0}^{\vec{x}} dA(\vec{x}_1)\vec{f}^{(3)}(\vec{x}_0) + \vec{f}^{(4)}(\vec{x}_0) \nonumber
% \\
% &= A(\vec{x})\int_{\vec{x}_0}^{\vec{x}} dA(\vec{x}_1)\vec{f}^{(2)}(\vec{x}_1)-\int_{\vec{x}_0}^{\vec{x}}A(\vec{x}_1)dA(\vec{x}_1)\vec{f}^{(2)}(\vec{x}_1) \nonumber
% \\
% &\quad + \int_{\vec{x}_0}^{\vec{x}} dA(\vec{x}_1)\vec{f}^{(3)}(\vec{x}_0) + \vec{f}^{(4)}(\vec{x}_0) \, .
\\
&= A(\vec{x}) \vec{f}^{(3)}(\vec{x}) - A(\vec{x}_0) \vec{f}^{(3)}(\vec{x}_0) - \int_{\vec{x}_0}^{\vec{x}}A(\vec{x}_1)dA(\vec{x}_1)\vec{f}^{(2)}(\vec{x}_1) + \vec{f}^{(4)}(\vec{x}_0) \, .
\end{align}

So far, we have discussed the solutions valid in the unphysical region. In practice, we need to do an analytic continuation to the physical region $s > 2m_h^2$. Up to weight 2, this can be simply done using the analytic expressions in Eq.~(\ref{eq:fij}) and (\ref{eq:weight_two}), with the branch choice according to $s \to s+i\delta$ and $m_t^2 \to m_t^2 - i\delta$. The treatment of the weight-3 and weight-4 parts is more tricky, since they are represented as one-fold integrals. We need to carefully deform the integration contour to avoid possible singularities.
For example, the integrals have a branch cut $1/\mu < -1$ on the real axis in the complex-$1/\mu$ plane, which corresponds to $s > 4m_t^2$. Suppose that we want to evaluate the integrals for a phase-space point at $1/\mu = \rho - i\delta$, with $\rho < -1$. We can integrate from the boundary point $1/\mu_0 = 0$ to the point $1/\mu$ along a half-circle below the real axis.
After the analytic continuation, we can numerically evaluate all the master integrals for topology E in the physical region. The results are shown in the next subsection.

\subsubsection{Numerical results for topology E}

\begin{figure}[t!]
\centering
\includegraphics[width=0.48\textwidth]{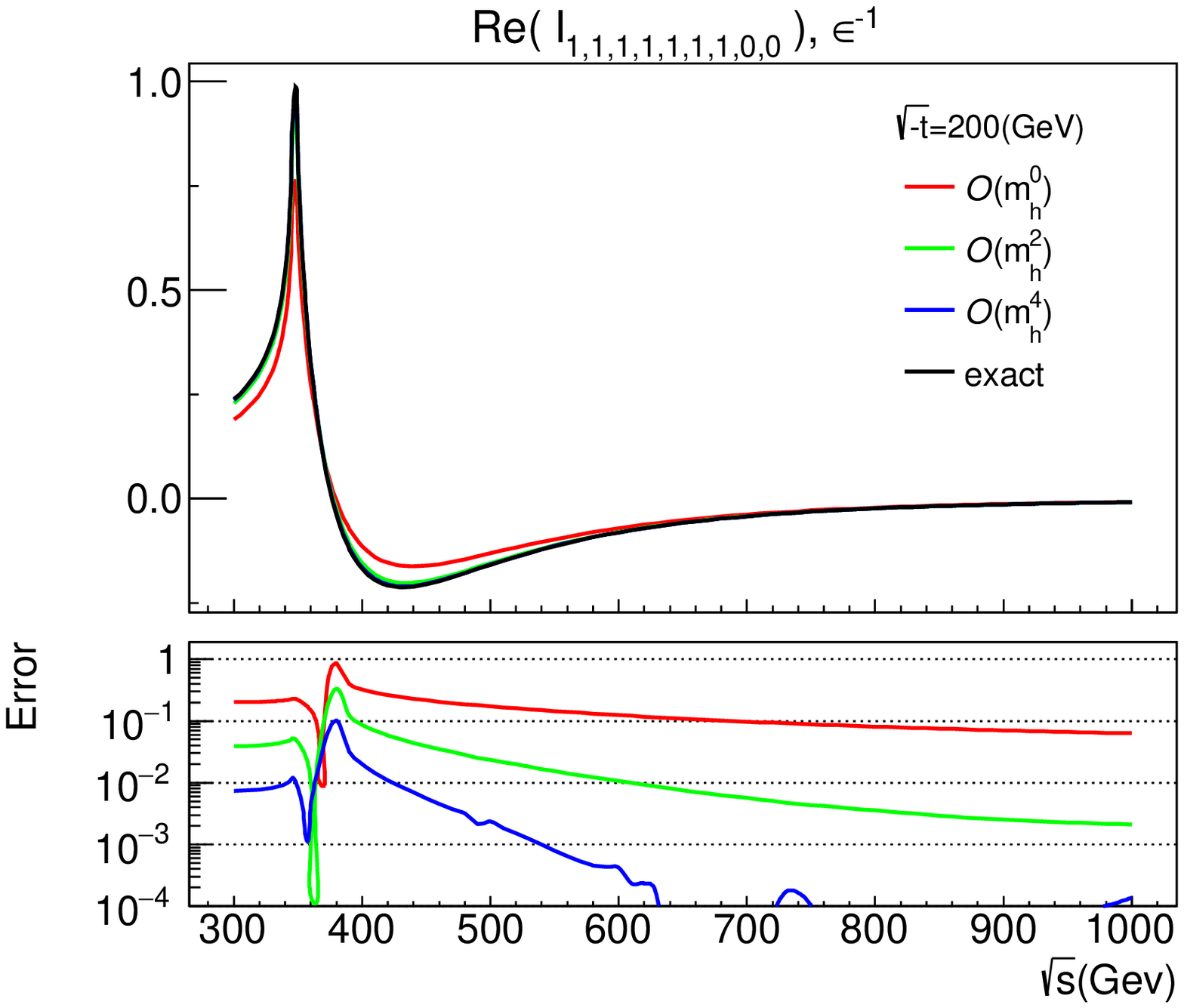}
\includegraphics[width=0.48\textwidth]{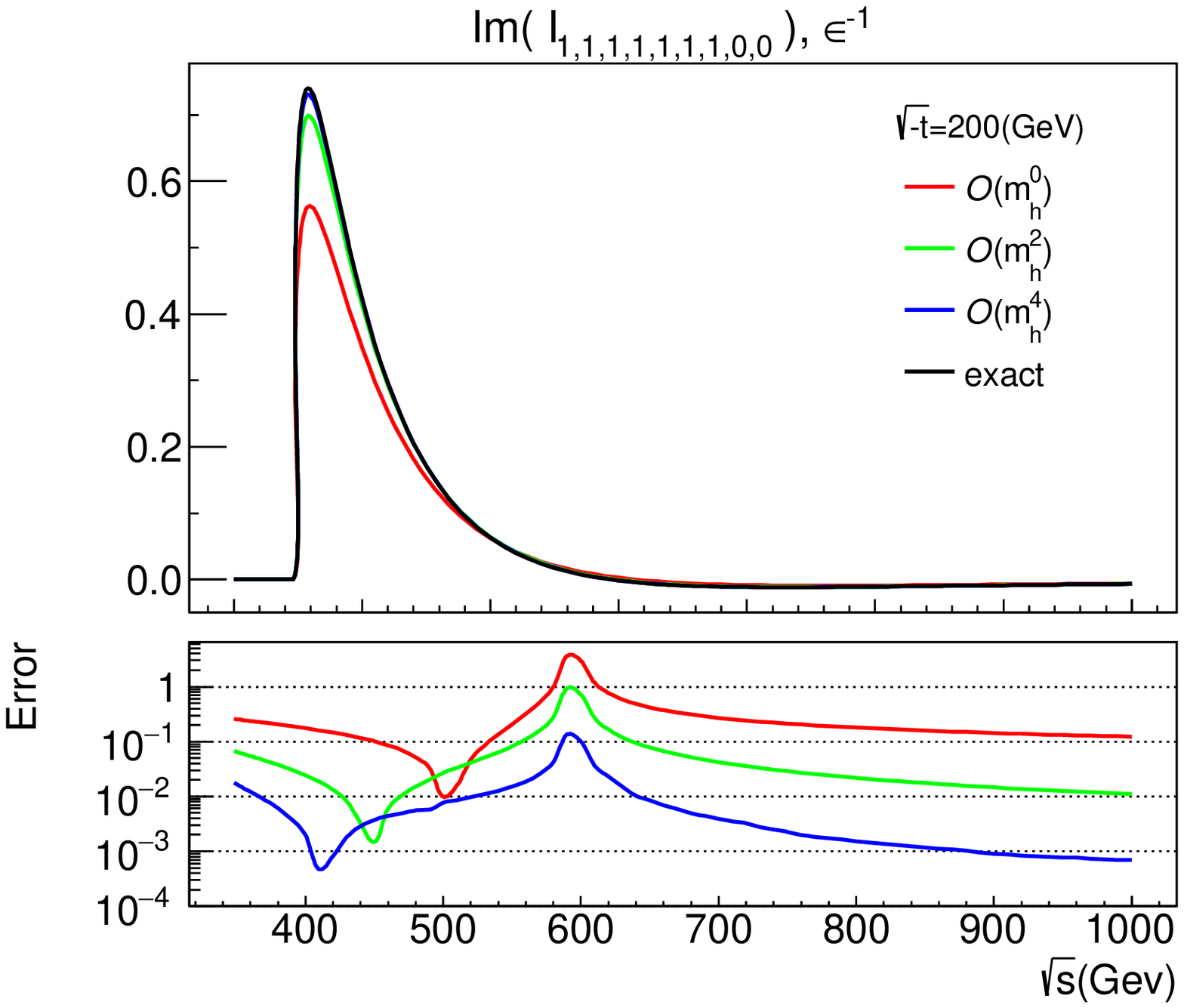}
\\
\vspace{-10ex}
\includegraphics[width=0.48\textwidth]{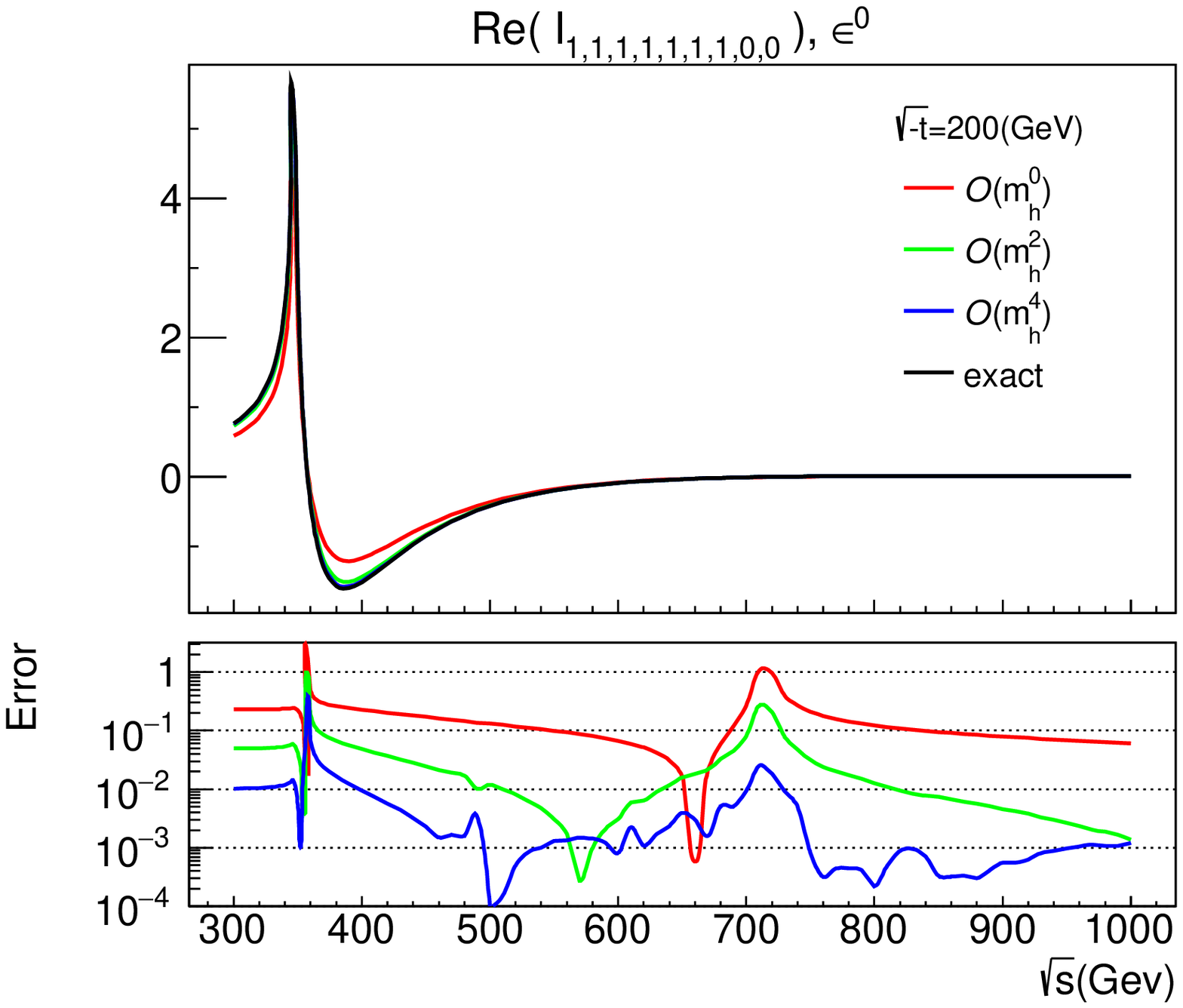}
\includegraphics[width=0.48\textwidth]{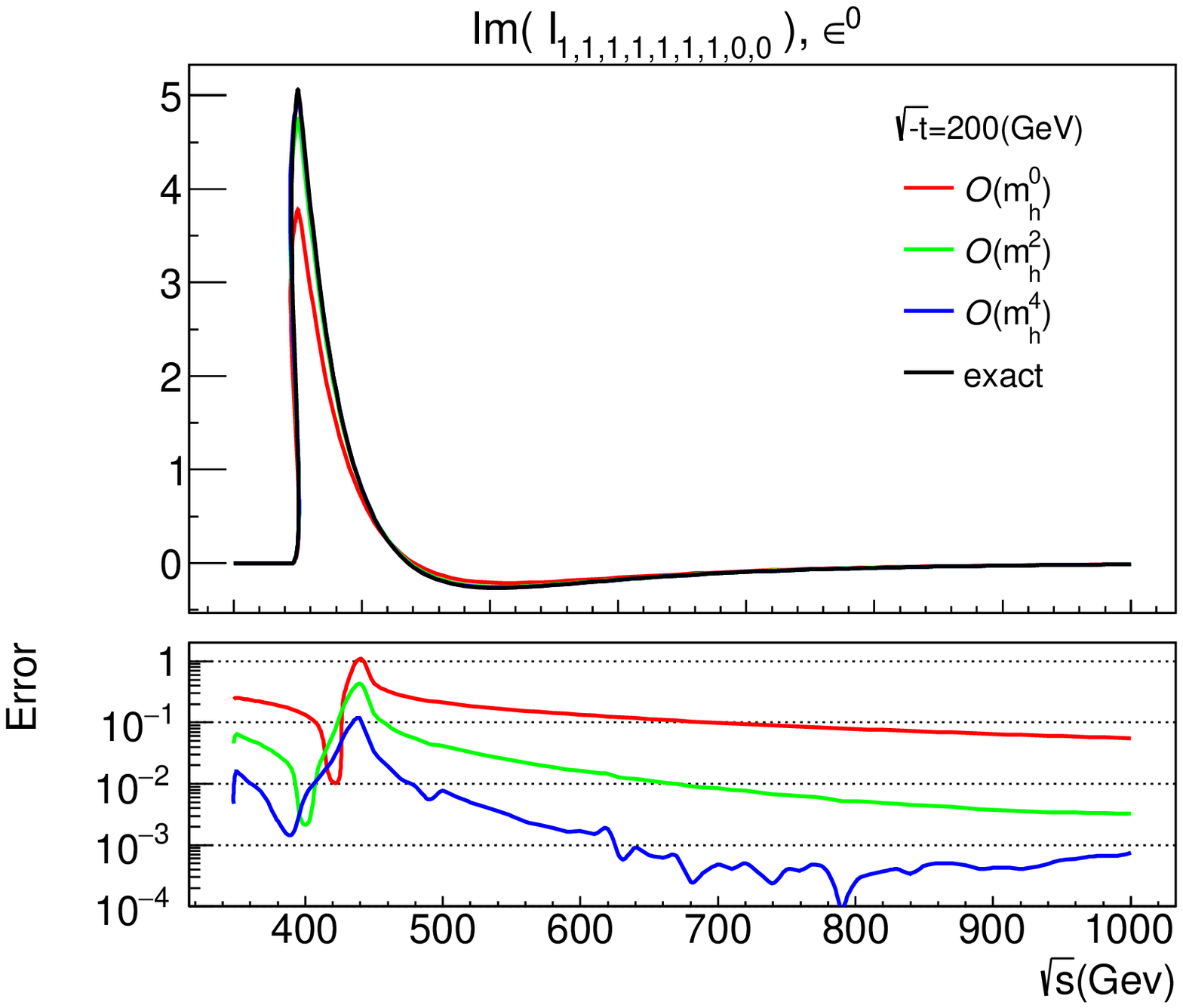}
\vspace{-12ex}
\caption{The real part (left two plots) and the imaginary part (right two plots) of the order $\epsilon^{-1}$ and $\epsilon^0$ coefficient for the two-loop integral $I_{1,1,1,1,1,1,1,0,0}$ in topology E as a function of $\sqrt{s}$ with $t=-(\unit{200}{\GeV})^2$. The integral has been multiplied by $m_t^6$ to make it dimensionless.}
\label{fig:TopoEs}
\end{figure}

In this subsection, we perform a numerical study of the 7-propagator two-loop integral $I_{1,1,1,1,1,1,1,0,0}$ in topology E. The purpose is to check how well the small-$m_h$ expansion can approximate the exact result. We calculate the exact result using the method of sector decomposition implemented in \texttt{pySecDec} \cite{Borowka:2017idc}. We perform the small-$m_h$ expansion up to order $m_h^4$, which can be extended to higher powers of $m_h$ straightforwardly.

As in the one-loop case, we first fix $\sqrt{-t} = \unit{200}{\GeV}$ and show the value of the integral as a function of $\sqrt{s}$ in Figure~\ref{fig:TopoEs}. The upper two plots show the coefficient of $\epsilon^{-1}$, and the lower two plots show the coefficient at $\epsilon^0$. We observe similar behaviors as the one-loop case: the small-$m_h$ expansion provides a good overall approximation to the exact result in the whole range of $\sqrt{s}$, from the threshold region $\sqrt{s} \gtrsim 2m_h$, to the $t\bar{t}$ threshold $\sqrt{s} \sim 2m_t$, and to the high energy regime $\sqrt{s} \gg 2m_t$. There are exceptional values of $\sqrt{s}$ where the relative errors grow, which is due to that the value of the integral is close to zero. One should however not be worried since in these phase space points, this integral is not expected to be the dominant contribution. Similar behaviors have been observed in the one-loop case, as was shown in Figure~\ref{fig:I1111} and \ref{fig:LOcs}. Even if there is a concern, one could easily add the order $m_h^6$ terms which will further improve the accuracy of the approximation.

\begin{figure}[t!]
\centering
\includegraphics[width=0.48\textwidth]{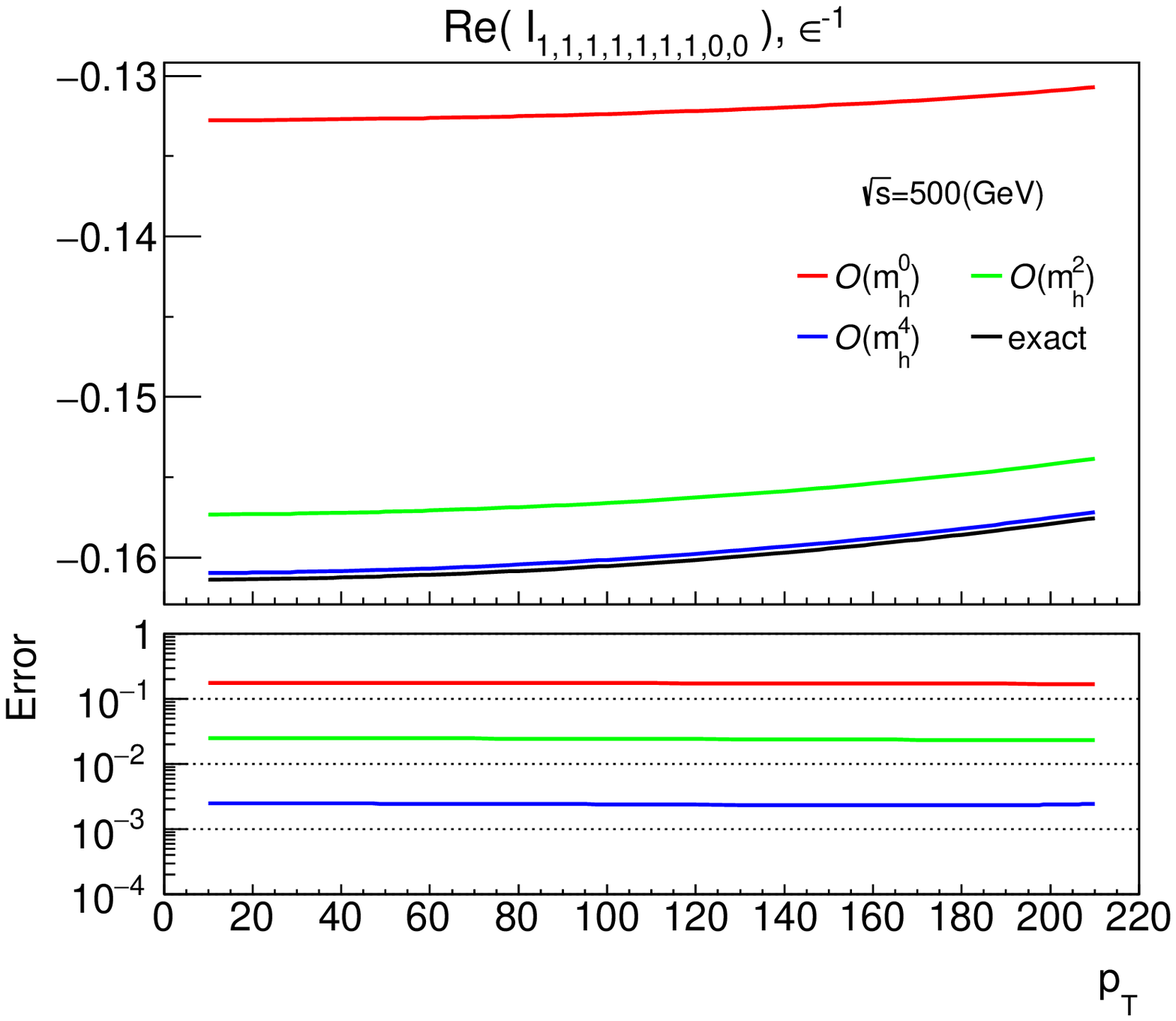} 
\includegraphics[width=0.48\textwidth]{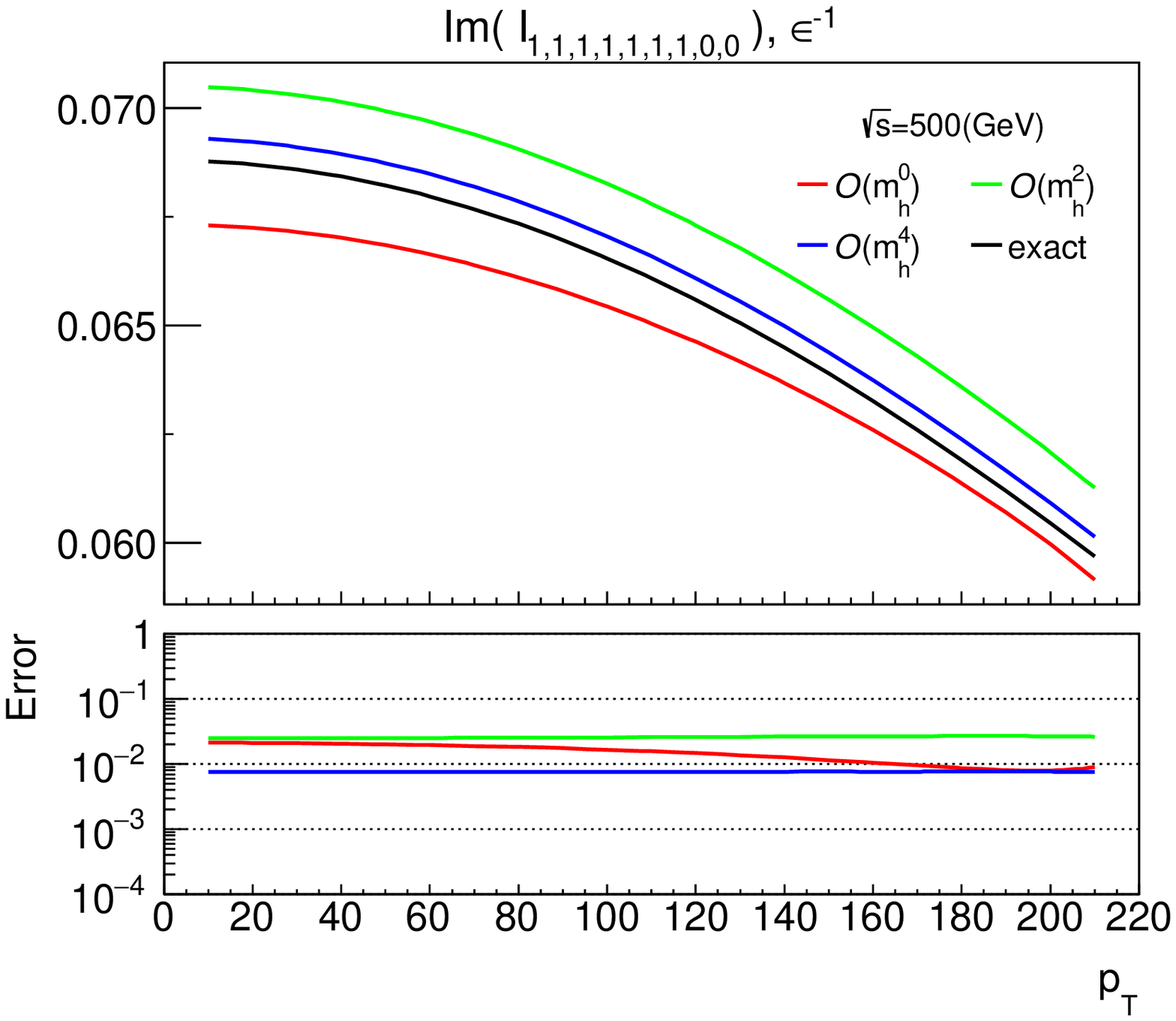}
\\
\vspace{-10ex}
\includegraphics[width=0.48\textwidth]{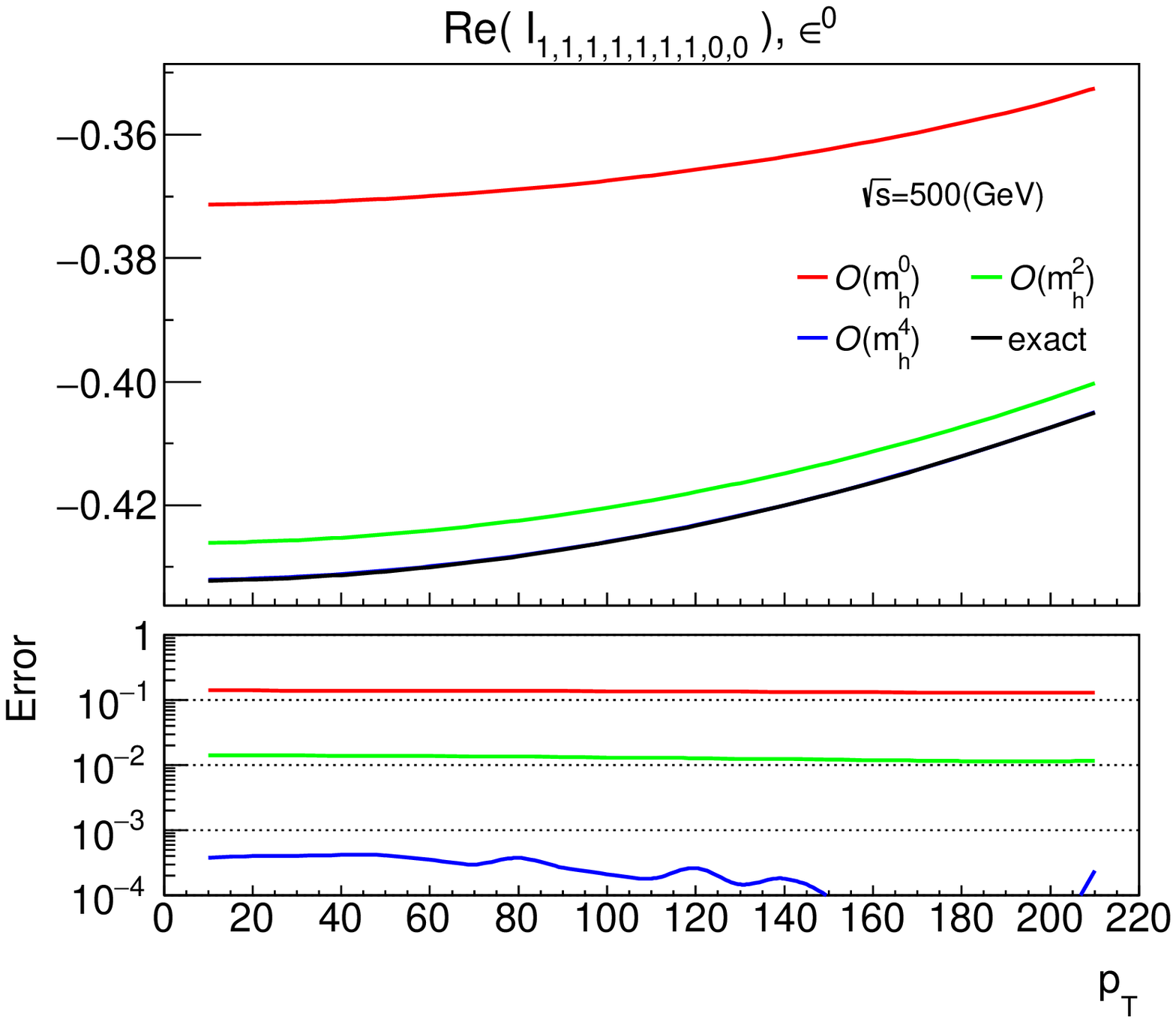} 
\includegraphics[width=0.48\textwidth]{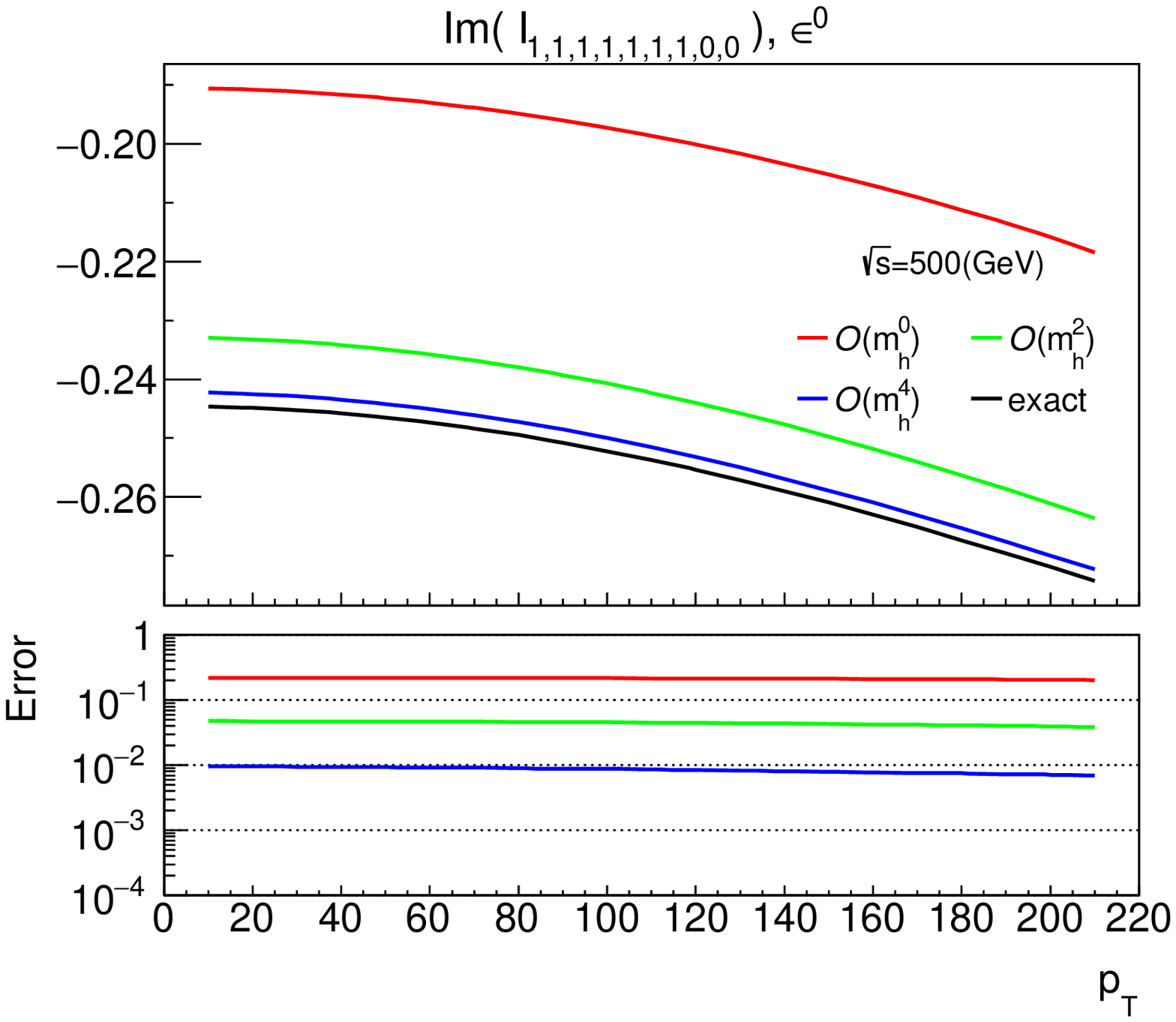}
\vspace{-12ex}
\caption{The real part (left two plots) and the imaginary part (right two plots) of the order $\epsilon^{-1}$ and $\epsilon^0$ coefficient of the two-loop integral $I_{1,1,1,1,1,1,1,0,0}$ in topology E as a function of the Higgs boson transverse momentum $p_T$ with $\sqrt{s}=\unit{500}{\GeV}$. The integral has been multiplied by $m_t^6$ to make it dimensionless.}
\label{fig:TopoEpT500}
\end{figure}

\begin{figure}[t!]
\centering
\includegraphics[width=0.48\textwidth]{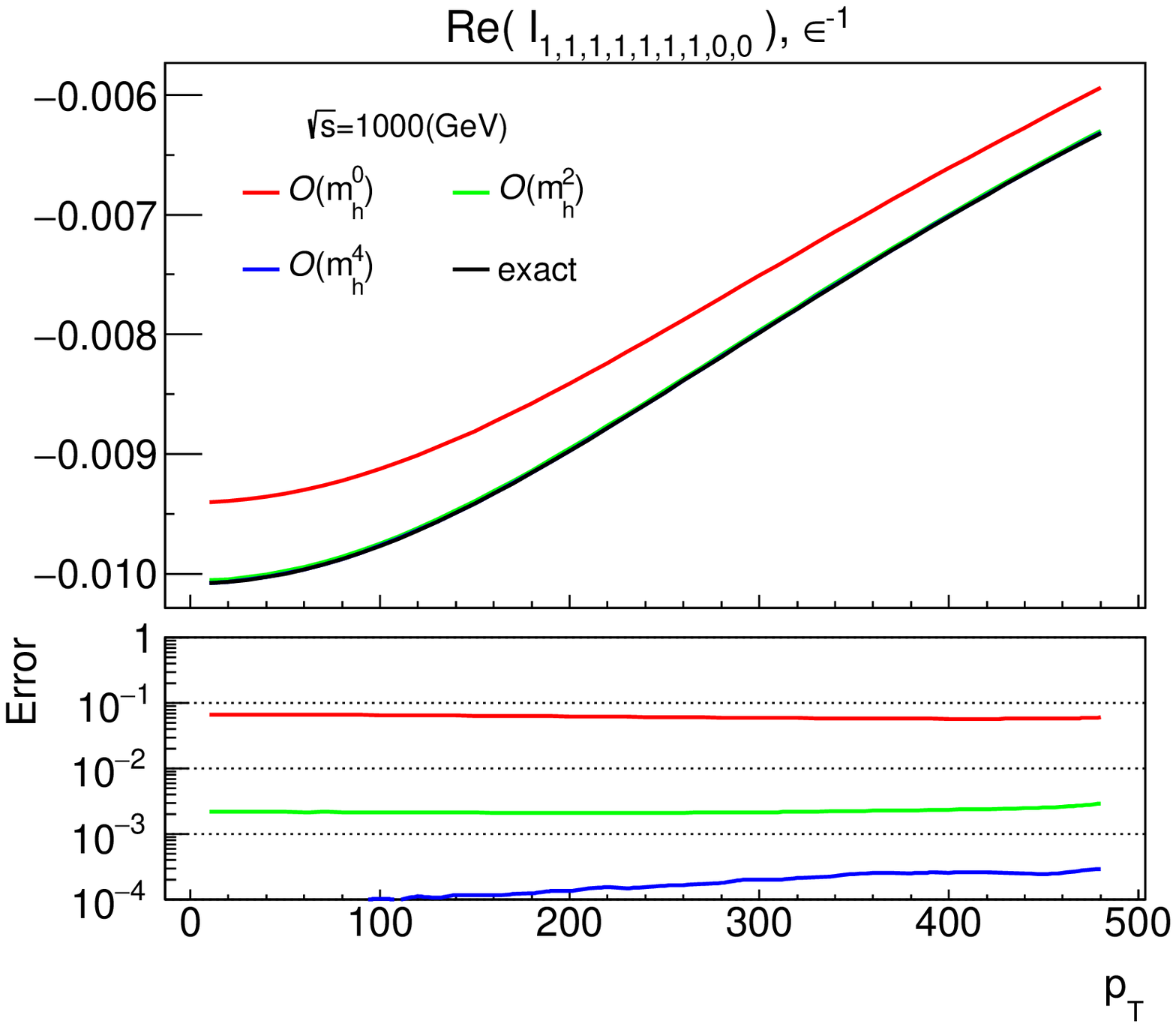} 
\includegraphics[width=0.48\textwidth]{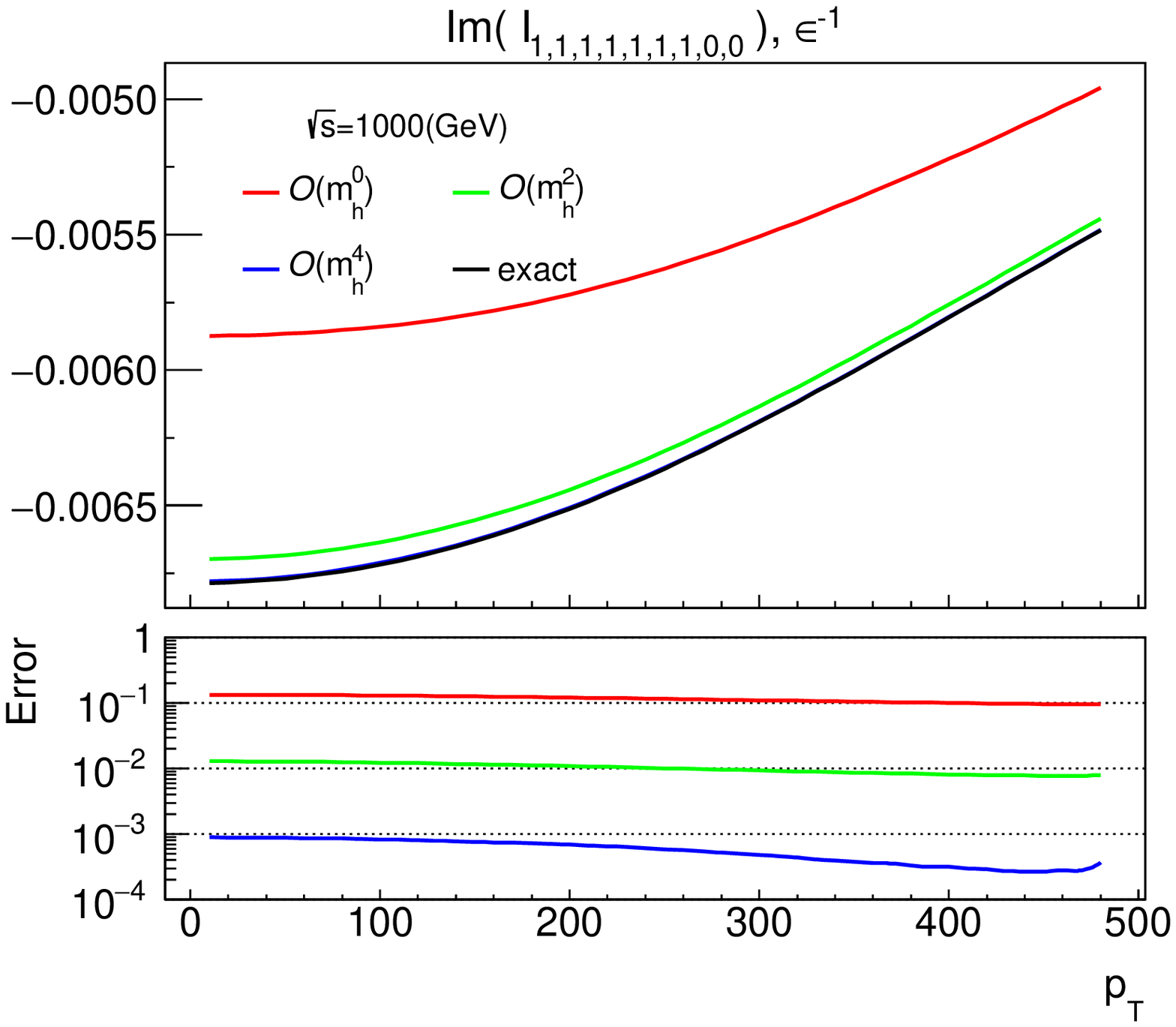}
\\
\vspace{-10ex}
\includegraphics[width=0.48\textwidth]{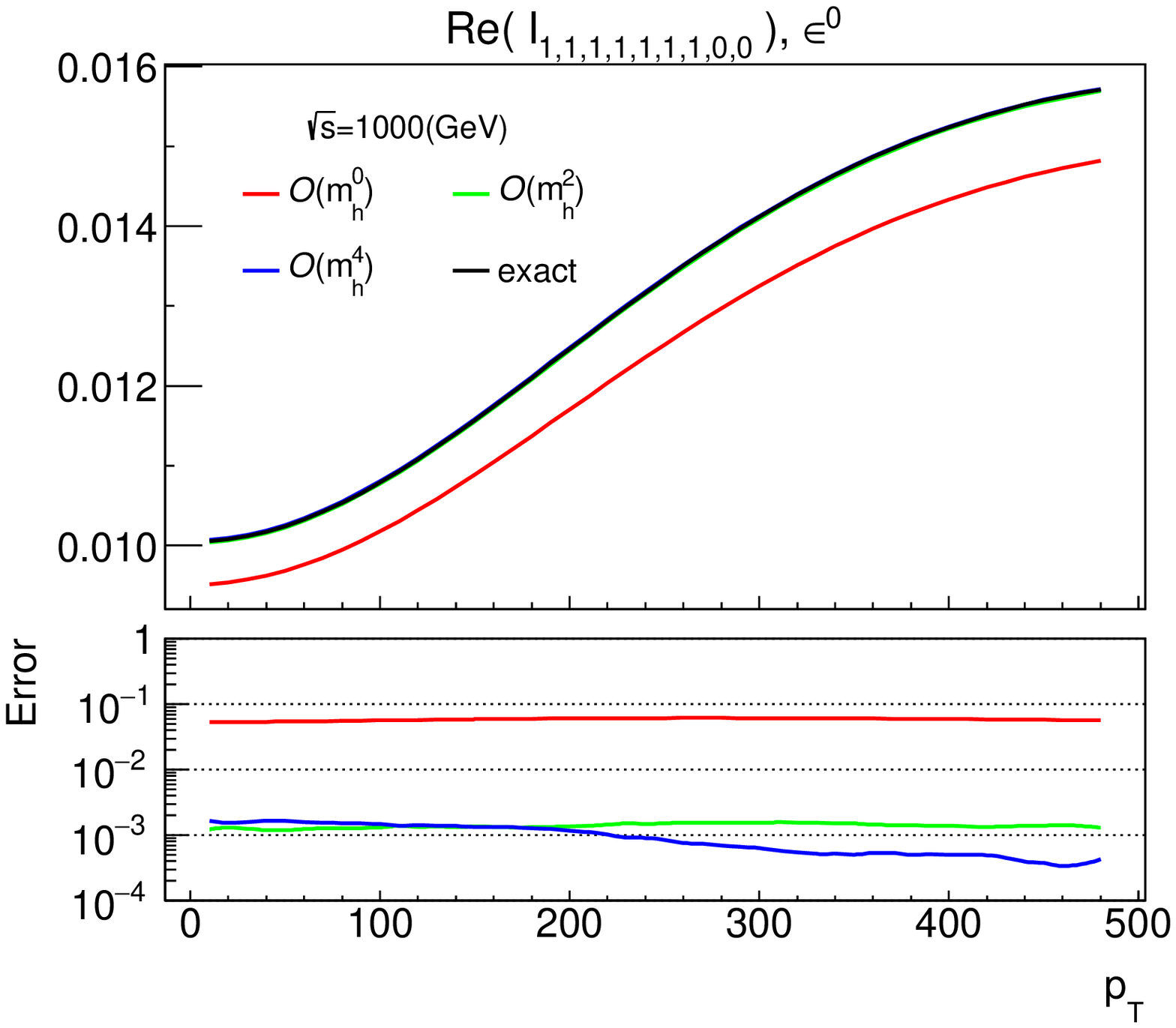} 
\includegraphics[width=0.48\textwidth]{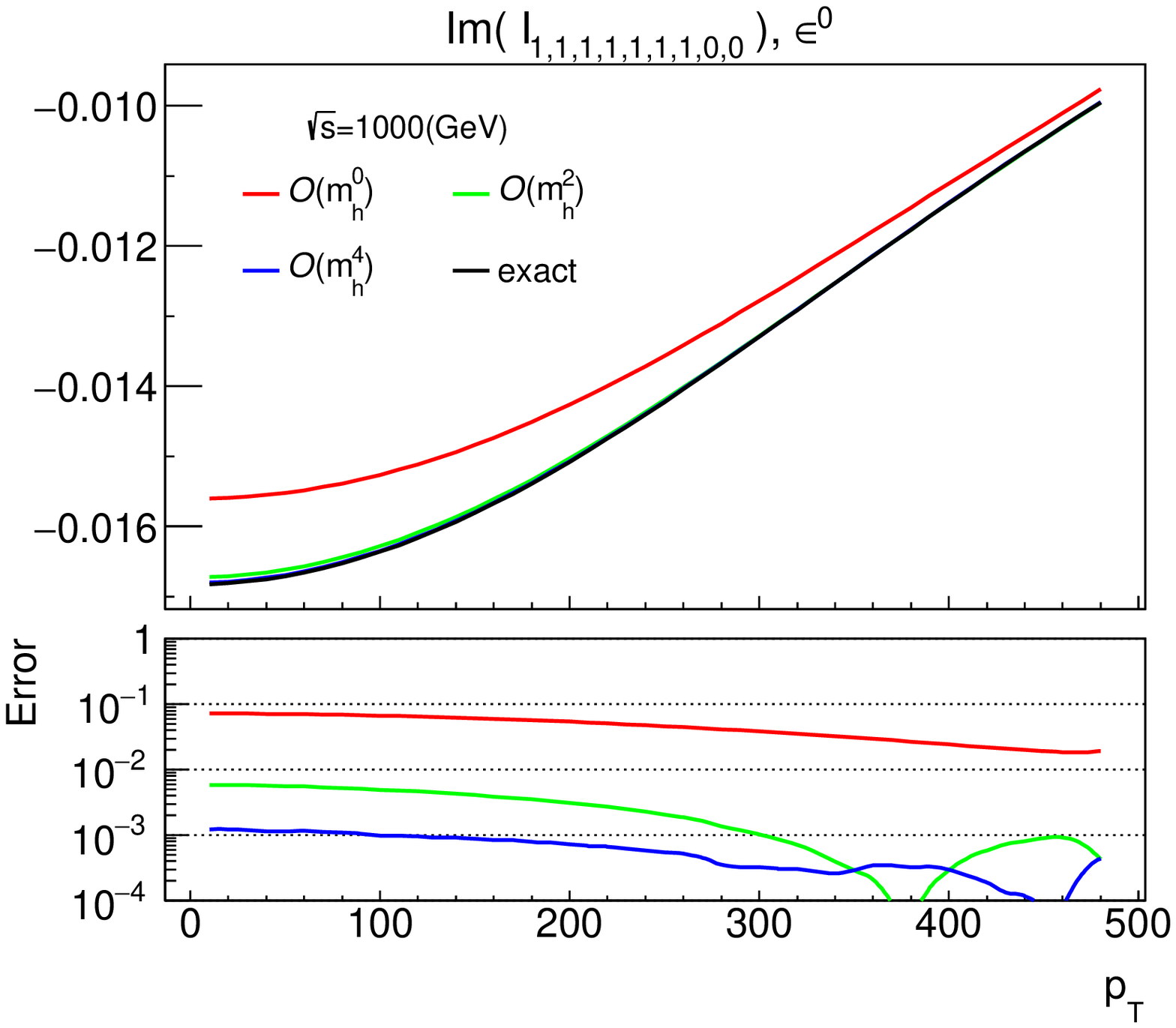}
\vspace{-12ex}
\caption{The real part (left two plots) and the imaginary part (right two plots) of the order $\epsilon^{-1}$ and $\epsilon^0$ coefficient of the two-loop integral $I_{1,1,1,1,1,1,1,0,0}$ in topology E as a function of the Higgs boson transverse momentum $p_T$ with $\sqrt{s}=\unit{1000}{\GeV}$. The integral has been multiplied by $m_t^6$ to make it dimensionless.}
\label{fig:TopoEpT1000}
\end{figure}

We further investigate the behavior of our approximation as a function of the transverse momentum $p_T$ of the Higgs boson. The invariant $t$ is related to $p_T$ by
\begin{equation}
t = \frac{2m_h^2-s \pm \sqrt{s^2-4m_h^2s-4p_T^2s}}{2} \, ,
\end{equation}
where the $\pm$ sign corresponds to the forward and backward scatterings, respectively. For convenience, we only show the results with the $+$ sign in the following. We take two typical values of the partonic center-of-mass energy: $\sqrt{s}=\unit{500}{\GeV}$ which is in the bulk region of the partonic cross section, and $\sqrt{s}=\unit{1000}{\GeV}$ which is in the high energy region. The corresponding numerical results are shown in Figure~\ref{fig:TopoEpT500} and \ref{fig:TopoEpT1000}, respectively. At $\sqrt{s}=\unit{500}{\GeV}$, we find that the approximation at order $m_h^4$ works rather well for the real part of the integral, with per-mille accuracy in the whole range of $p_T$. For the imaginary part, the accuracy is about 1\%, and if one needs to have a better approximation, the order $m_h^6$ terms should be added. When the center-of-mass energy goes higher, at $\sqrt{s}=\unit{1000}{\GeV}$, the quality of the approximation becomes better, with per-mille accuracy in all situations. This can be expected since in the high energy region all the scales are much larger than $m_h$.

We stress that although in this subsection we only studied the behavior of a single integral, similar behavior is expected for the full amplitude. This has been verified at the one-loop level. At the two-loop level, this can only be done with the results for topology F, which is the subject of the next subsection.

Finally, we emphasize that due to the analytic nature, the evaluation of the integrals up to weight 2 is extremely fast. The weight-3 and weight-4 parts involve one-fold integrals to be performed. We have carried out the integration using \texttt{Mathematica} on a desktop computer with 6 cores, without too much optimization. We have checked that to evaluate \textbf{all} the master integrals $\vec{f}$ in topology E (which can be used to construct all the integrals $I_{\{a_i\}}$ by simple arithmetic operations) for one phase-space point, it takes about 20 seconds with 6 threads. We believe that by using a dedicated \texttt{C++} code and by performing a bit of optimization, the time can be significantly shortened.
For comparison, to evaluate just \textbf{one} master integral $I_{1,1,1,1,1,1,1,0,0}$ with \texttt{pySecDec} on the same computer, it takes about 25 minutes with 12 threads.

\subsection{Towards a solution for topology F}
\label{sec:topologyF}

\begin{figure}[t!]
\centering
\includegraphics[width=6cm,height=3cm]{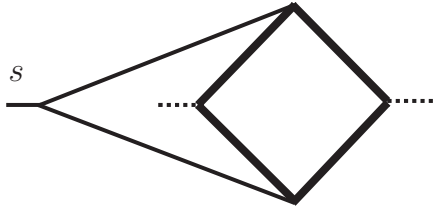}
\vspace{-1ex}
\caption{Elliptic sub-topology for topology F. The thick lines represent massive propagators (top quarks), while the thin lines represent massless propagators (gluons). The dashed external legs are light-like, while the solid external leg is massive.}
\label{fig:sub-topology}
\end{figure}

Topology F is the most difficult one as the differential equations for the master integrals cannot be transformed into a canonical form. The first place where this shows up is the 6-propagator sub-topology depicted in Figure~\ref{fig:sub-topology} (which has been discussed in \cite{vonManteuffel:2017hms}). We denote the 6-propagator master integrals as $\vec{f}(\mu,\epsilon)$, and collect the master integrals with fewer propagators in $\vec{g}(\mu,\epsilon)$, where $\mu=-4m_t^2/s$ as before. Then the differential equation satisfied by $\vec{f}(\mu,\epsilon)$ can be written as
\begin{align}
\frac{d}{d\mu} \vec{f}(\mu,\epsilon) = \big( \epsilon A(\mu) + B(\mu) \big) \vec{f}(\mu,\epsilon)+ C(\mu,\epsilon) \vec{g}(\mu,\epsilon) \, .
\end{align}
The transformation to a canonical form amounts to get rid of the order $\epsilon^0$ coefficient matrix $B(\mu)$ in the above equation. However, for the topology in Figure~\ref{fig:sub-topology}, we find that differential equations for the two top-level master integrals
\begin{align}
\vec{f}(\mu,\epsilon) = \left\{ \frac{\epsilon (1+4\epsilon)}{\mu^2} \tilde{I}_{1,0,1,1,1,1,1,0,0}, \, \frac{\epsilon}{\mu^2} \tilde{I}_{1,0,1,2,1,1,1,0,0} \right\}
\end{align}
involves the non-diagonal coefficient matrix at order $\epsilon^0$:
\begin{align}
B(\mu) =
\begin{pmatrix}
0 & \frac{4}{\mu}
\\ 
\frac{1}{4(1-4\mu)} & \frac{4}{1-4\mu}
\end{pmatrix}
\end{align}
which cannot be transformed away. We call this situation as a ``two-coupled'' system of differential equations. In this case the solution necessarily involves elliptic integrals. To see that, we turn the system of two first-order differential equations into a second-order differential equation for $f_1^{(n)}(\mu)$
\begin{align}
\frac{d^2}{d\mu^2} f_1^{(n)}(\mu) + a(\mu) \frac{d}{d\mu} f_1^{(n)}(\mu) + b(\mu) f_1^{(n)}(\mu) = c_1^{(n)}(\mu) \, ,
\end{align}
where the rational functions $a(\mu)$ and $b(\mu)$ are related to the matrix $B(\mu)$, and the function $c_1^{(n)}(\mu)$ depends on $A(\mu)$, $C(\mu,\epsilon)$ and $\vec{g}(\mu,\epsilon)$. The homogeneous part of the above equation (i.e., with $c_1^{(n)}(\mu)$ absent) can be solved in terms of elliptic integrals, upon which the inhomogeneous part of the solution can be added.

We now turn to the integrals with 7 propagators in topology F, which to our knowledge were not discussed in the literature. There are 4 top-level master integrals in this case, and we are facing a four-coupled system of differential equations to begin with. In order to reduce the system to smaller blocks, we employ the method of \cite{Adams:2017tga}. Briefly speaking, the method goes as follows. We consider an $N$-coupled system of differential equations with respect to a single kinematic variable $x$ (the extension to multiple variables is straightforward). We start from one of the master integrals $I$, and study the derivatives $d^kI/dx^k$. In general, not all the derivatives are linearly independent, and we can find the maximal number of independent derivatives $I$, $dI/dx$, \ldots, $d^{(r-1)}I/dx^{(r-1)}$. If $r < N$, we can choose these $r$ integrals as new master integrals such that they are decoupled from the remaining $N-r$ master integrals.

Applying the above method, we find the basis of the top-level integrals in topology F
\begin{align}
\vec{h}(\vec{x},\epsilon) &= \bigg\{ \frac{ \tilde{I}_{1,1,1,1,1,1,1,0,-1}}{\mu^2} , \, \frac{2(1-3\mu) \, \tilde{I}_{1,1,1,1,1,1,1,0,-1}}{\mu^2(1-4\mu)} + \frac{\nu (\tilde{I}_{1,1,1,1,1,1,1,-1,-1}-\tilde{I}_{1,1,1,1,1,1,1,0,-2})}{2\mu(1-4\mu)(2\mu+\nu)} , \nonumber 
\\
&\hspace{-2em} \frac{\mu \nu \tilde{I}_{1,1,1,1,1,1,1,-1,-1} + 2(2\mu+\nu) \tilde{I}_{1,1,1,1,1,1,1,-1,-1}}{\mu^2\nu} , \, \frac{\sqrt{\mu+\nu}\sqrt{\mu+\nu+\mu^2} \, \tilde{I}_{1,1,1,1,1,1,1,0,0}}{\mu^2\nu^2} \bigg\} \, .
\end{align}
The differential equation takes the form
\begin{align}
d\vec{h}(\vec{x},\epsilon) = \big( \epsilon dA(\vec{x}) + dB(\vec{x}) \big) \vec{h}(\vec{x},\epsilon) + \cdots \, ,
\end{align}
where the order $\epsilon^0$ coefficient matrix $dB(\vec{x})$ takes the illustrative form
\begin{align}
dB(\vec{x}) = 
\begin{pmatrix}
* & * & 0 & 0
\\
* & * & 0 & 0
\\
* & * & 0 & 0
\\
0 & 0 & 0 & 0
\end{pmatrix}
\, ,
\end{align}
in which ``$*$'' denotes non-zero entries. It is then clear that the 7-propagator master integrals can also be solved in terms of elliptic integrals. That said, the solutions are still rather complicated, and we leave them for future works.

\section{Conclusion and outlook}
\label{sec:conclusion}

In this paper we propose a new method to evaluate loop integrals where the masses of the internal particles are larger than the external particles. This can be applied to the pair production process and the associated production processes of the Higgs boson, which are mainly mediated by top quark loops. Our method amounts to perform a Taylor expansion in terms of the small masses of external particles. The coefficients of the expansion are written in terms of loop integrals with fewer mass scales than the original integrals, and are therefore easier to evaluate. The main difference between our method and other expansion methods lies in the fact that the validity of our expansion is not restricted to a special phase space region. Instead, our method provides a systematically improvable approximation in the entire phase space. Our expansion works particularly well for the high energy tails of kinematic distributions where many other expansions cease to be valid.

We demonstrate our method using Higgs boson pair production as an example. At the leading order (one loop), we compare the approximate and the exact results both at the level of a single master integral and at the level of differential cross sections. We find that our method leads to rather good approximations in both cases. At the next-to-leading order (two-loop), we expand the amplitude and classify the resulting loop integrals into 2 planar topologies and 2 non-planar topologies. We reduce these integrals to master integrals using IBP reduction, and derive differential equations satisfied by the master integrals. We find that the equations for the 2 planar topologies and the non-planar topology E can be casted into the canonical form, they can be solved in terms of Chen iterated integrals. The solutions up to weight 2 can be written in terms of logarithms and dilogarithms, while the weight 3 and weight 4 parts of the solutions are given as one-fold integrals. We present numeric results for an integral appearing in the original amplitude (with non-zero external masses), comparing the exact values from sector decomposition and the approximate values from our expansion. We observe similar behaviors as in the one-loop case, that our expansion up to order $m_h^4$ leads to good approximations in the whole phase space, which can still be further improved by incorporating terms suppressed by more powers of $m_h$.

To construct the approximation to the full two-loop amplitude, we still need to calculate the master integrals in the other non-planar topology (topology F). The differential equations for the topology cannot be transformed into a canonical form. We reduce the system of differential equations into smaller blocks, and find that they can be solved in terms of elliptic integrals. The full solution and the numeric study for this topology will be presented in another work. Combining our efficient method to compute the two-loop amplitude with an infrared subtraction method for the real emission corrections, we expect to have a fast and reliable tool to calculate the differential cross sections for Higgs boson pair production. This will be useful for phenomenological studies and for the extraction of the Higgs self-coupling from future experimental data.

Our method is not restricted to double Higgs production. It can be applied whenever the internal masses in the loop are larger than the external masses, without the restrictions put on the Mandelstam variables. We hope to see applications of our method to further processes such as $H+j$ production and $H+Z$ production.

\section*{Acknowledgements}
We would like to thank Lorenzo Tancredi for useful discussions. This work was supported in part by the National Natural Science Foundation of China under Grant No. 11575004 and 11635001.

\appendix

\section{Appendix}

In this appendix, we provide the canonical basis for topology E, where the integrals $\tilde{I}_{\{a_i\}}$ are defined in Eqs.~(\ref{eq:twoloopint})--(\ref{eq:propagators}). The general procedure to derive the canonical basis is described in Section~\ref{sec:topologyE}. The canonical basis is expressed in terms of the pre-canonical integrals show in Figure~\ref{fig:pre-canonical} and is given by
\begin{align*}
f_{1}&=\epsilon ^2 \tilde{I}_{0,0,0,0,0,2,2,0,0} \, ,\\[-0.3cm]
f_{2}&=- \frac{4 \epsilon ^2 \sqrt{\mu+1} }{\mu} \tilde{I}_{0,0,0,1,0,2,2,0,0} \, ,\\[-0.3cm]
f_{3}&=\frac{\epsilon ^2}{\mu} \tilde{I}_{0,1,0,2,0,0,2,0,0} \, ,\\[-0.3cm]
f_{4}&=-\frac{4 \epsilon ^2 \sqrt{\mu+1} }{\mu}( \tilde{I}_{0,1,0,2,0,0,2,0,0}+2\tilde{I}_{0,2,0,1,0,0,2,0,0}) \, ,\\[-0.3cm]
f_{5}&=\frac{\epsilon ^2 \tilde{I}_{0,1,2,0,2,0,0,0,0}}{\nu} \, ,\\[-0.3cm]
f_{6}&=-\frac{4\epsilon ^2 \sqrt{\nu+1}  }{\nu}(\tilde{I}_{0,1,2,0,2,0,0,0,0}+2 \tilde{I}_{0,2,2,0,1,0,0,0,0}) \, ,\\[-0.3cm]
f_{7}&=\frac{\epsilon ^2  (\mu+\nu)}{\mu \nu}\tilde{I}_{1,0,0,0,2,0,2,0,0} \, ,\\[-0.3cm]
f_{8}&=\frac{\epsilon ^2 \sqrt{\mu+\nu} \sqrt{\mu+\nu-\mu \nu}}{\mu \nu}( \tilde{I}_{1,0,0,0,2,0,2,0,0}+2\tilde{I}_{2,0,0,0,2,0,1,0,0}) \, ,\\[-0.3cm]
f_{9}&=\frac{\epsilon ^3}{\mu} \tilde{I}_{0,0,0,1,1,1,2,0,0} \, ,\\[-0.3cm]
f_{10}&=\frac{\epsilon ^3}{\mu} \tilde{I}_{0,1,0,1,1,0,2,0,0} \, ,\\[-0.3cm]
f_{11}&=\frac{\epsilon ^3}{\nu}\tilde{I}_{0,1,1,0,2,0,1,0,0} \, ,\\[-0.3cm]
f_{12}&=\frac{\epsilon ^3 (\mu+\nu)}{\mu \nu}\tilde{I}_{1,0,0,0,1,1,2,0,0} \, ,\\[-0.3cm]
f_{13}&=\frac{\epsilon ^2}{\mu} \tilde{I}_{0,1,3,1,0,1,0,0,0} \, ,\\[-0.3cm]
f_{14}&=\frac{\epsilon ^3}{\mu} \tilde{I}_{0,1,2,1,0,1,0,0,0} \, ,\\[-0.3cm]
f_{15}&=\frac{ \epsilon ^2 \sqrt{\mu+1}  }{ \mu}\left(\frac{3}{2}\epsilon \tilde{I}_{0,2,1,1,0,1,0,0,0}+\tilde{I}_{0,1,2,2,0,1,0,0,0}-2\tilde{I}_{0,1,3,1,0,1,0,0,0}\right)\, ,\\[-0.3cm]
f_{16}&=\frac{\epsilon ^4}{\nu} \tilde{I}_{0,1,1,0,1,1,1,0,0} \, ,\\[-0.3cm]
f_{17}&=\frac{\epsilon ^4 (\mu+\nu)}{\mu \nu} \tilde{I}_{1,0,1,1,1,0,1,0,0} \, ,\\[-0.3cm]
f_{18}&=\frac{ \epsilon ^3 \sqrt{\mu+1}}{\mu^2}\tilde{I}_{0,1,1,2,0,1,1,0,0} \, ,\\[-0.3cm]
f_{19}&=\frac{ \epsilon ^3(\mu+2) }{\mu^2}\tilde{I}_{0,1,1,2,0,1,1,0,0}+\frac{\epsilon ^3 }{\mu}\tilde{I}_{0,1,1,1,0,2,1,0,0} \, ,\\[-0.3cm]
f_{20}&=\frac{\epsilon ^4  (\mu+\nu)}{\mu \nu}\tilde{I}_{0,1,1,1,1,0,1,0,0} \, ,\\[-0.3cm]
f_{21}&=\frac{\epsilon ^3  \sqrt{\mu+\nu+1}}{\mu \nu}\tilde{I}_{0,2,1,1,1,0,1,0,0} \, ,\\[-0.3cm]
f_{22}&=\frac{\epsilon ^3  (\mu+\nu)}{\mu \nu}(\tilde{I}_{0,1,1,1,1,0,2,0,0}+\tilde{I}_{0,1,2,1,1,0,1,0,0}) \, ,\\[-0.3cm]
f_{23}&=\frac{\epsilon ^4}{\nu} \tilde{I}_{1,0,1,0,1,1,1,0,0} \, , \\[-0.3cm]
f_{24}&=\frac{\epsilon ^3  \sqrt{\mu+\nu} \sqrt{\mu^2+\mu+\nu}}{\mu^2 \nu}\tilde{I}_{2,0,1,0,1,1,1,0,0} \, ,\\[-0.3cm]
f_{25}&=\frac{\epsilon ^3 }{\nu}(\tilde{I}_{1,0,1,0,1,1,2,0,0}+\tilde{I}_{1,0,2,0,1,1,1,0,0}) \, , \\[-0.3cm]
f_{26}&=\frac{\epsilon ^3  \sqrt{16 \mu+(\nu+4)^2}}{\mu \nu}\tilde{I}_{0,1,2,1,1,1,0,0,0} \, ,\\[-0.3cm]
f_{27}&=\frac{\epsilon ^3  \sqrt{\mu+\nu+1}}{\mu \nu}(\epsilon \tilde{I}_{0,1,2,1,1,1,0,0,0}+\tilde{I}_{0,1,3,1,1,1,0,0,0})\, , \\[-0.3cm]
f_{28}&=\frac{\epsilon ^2}{\mu}( \tilde{I}_{0,1,2,1,1,1,0,-1,0}+ \tilde{I}_{0,1,2,1,1,1,0,0,0})\, , \\[-0.3cm]
f_{29}&=\frac{\epsilon ^3 (\mu+\nu) \sqrt{\left[4-\mu\nu/(\mu+\nu)\right]^2+16\mu}}{\mu^2 \nu} \, \tilde{I}_{1,0,0,1,1,1,2,0,0}\,  ,\\[-0.3cm]
f_{30}&=\frac{\epsilon ^2 \sqrt{\mu+\nu} \sqrt{\mu^2+\mu+\nu}}{\mu^2 \nu}(\epsilon \tilde{I}_{1,0,0,1,1,1,2,0,0} +\tilde{I}_{1,0,0,1,1,1,3,0,0}) \, ,\\[-0.3cm]
f_{31}&=\frac{ \epsilon ^3(\mu+4)  }{\mu^2}\tilde{I}_{1,0,0,1,1,1,2,0,0}+\frac{\epsilon ^3 }{\mu}(\tilde{I}_{1,0,0,0,1,1,2,0,0}+\tilde{I}_{1,0,0,1,1,0,2,0,0}-\tilde{I}_{1,0,0,1,1,1,2,-1,0}) \, ,\\[-0.3cm]
f_{32}&=\frac{\epsilon ^4 }{\nu}\tilde{I}_{1,1,0,1,1,0,1,0,0}\, ,\\[-0.3cm]
f_{33}&=\frac{\epsilon ^3  \sqrt{\mu+\nu} \sqrt{\mu^2+\mu+\nu}}{\mu^2 \nu}\tilde{I}_{1,1,0,1,1,0,2,0,0} \, ,\\[-0.3cm]
f_{34}&=\frac{\epsilon ^3  \left(\mu^2+2 \mu+2 \nu\right)}{\mu^2 \nu}\tilde{I}_{1,1,0,1,1,0,2,0,0}+\frac{\epsilon ^3 }{\nu}(\tilde{I}_{1,1,0,1,2,0,1,0,0}+\tilde{I}_{1,1,0,2,1,0,1,0,0}) \, ,\\[-0.3cm]
f_{35}&=\frac{\epsilon ^4}{\mu} \tilde{I}_{1,1,1,0,1,0,1,0,0} \, ,\\[-0.3cm]
f_{36}&=\frac{\epsilon ^3 \sqrt{(\mu+\nu)} \sqrt{\nu^2+\mu+\nu}}{\mu \nu^2} \tilde{I}_{1,1,1,0,2,0,1,0,0} \, ,\\[-0.3cm]
f_{37}&=\frac{\epsilon ^3  (\nu^2+2 \mu+2\nu )}{\mu \nu^2}\tilde{I}_{1,1,1,0,2,0,1,0,0}+\frac{\epsilon ^3 }{\mu}(\tilde{I}_{1,1,1,0,1,0,2,0,0}+\tilde{I}_{1,1,2,0,1,0,1,0,0}) \, ,\\[-0.3cm]
f_{38}&=\frac{\epsilon ^4  (\mu+\nu)}{\mu \nu}\tilde{I}_{1,1,1,0,1,1,0,0,0}\, , \\[-0.3cm]
f_{39}&=\frac{\epsilon ^3  \sqrt{\mu+\nu+1}}{\mu \nu}\tilde{I}_{1,1,2,0,1,1,0,0,0} \, ,\\[-0.3cm]
f_{40}&=\frac{\epsilon ^3 (\mu+\nu+2)}{\mu \nu} \tilde{I}_{1,1,2,0,1,1,0,0,0}+\frac{\epsilon ^3  (\mu+\nu)}{\mu \nu}(\tilde{I}_{1,1,1,0,1,2,0,0,0}+ \tilde{I}_{1,1,1,0,2,1,0,0,0})\, , \\[-0.3cm]
f_{41}&=\frac{\epsilon ^4 }{\mu^2}\tilde{I}_{1,1,1,1,0,1,1,0,0} \, ,\\[-0.3cm]
f_{42}&=\frac{\epsilon ^4  \sqrt{\mu+\nu}}{\mu \nu}\tilde{I}_{1,1,1,0,1,1,1,0,0} \, ,\\[-0.3cm]
f_{43}&=\frac{\epsilon ^3  (\mu+\nu)}{4 \mu^2 \nu^2}(\nu \tilde{I}_{2,0,1,0,1,1,1,0,0}+\mu \tilde{I}_{1,1,1,0,2,0,1,0,0})+\frac{\epsilon ^3 }{4 \mu \nu}(\tilde{I}_{1,1,2,0,1,1,0,0,0}+4\tilde{I}_{1,1,2,0,1,1,1,0,0}) \, ,\\[-0.3cm]
f_{44}&=\frac{\epsilon ^4  \sqrt{\mu+\nu}}{\mu \nu}\tilde{I}_{1,1,1,1,1,0,1,0,0} \, ,\\[-0.3cm]
f_{45}&=-\frac{\epsilon ^3 (\mu+\nu)}{4 \mu^2 \nu}( \nu \tilde{I}_{1,1,0,1,1,0,2,0,0}-\mu \tilde{I}_{1,1,1,0,2,0,1,0,0}-4\mu  \tilde{I}_{1,1,1,1,2,0,1,0,0}-\mu \nu \tilde{I}_{0,2,1,1,1,0,1,0,0})\, , \\[-0.3cm]
f_{46}&=\frac{\epsilon ^4  \sqrt{\mu+\nu+1}}{\mu \nu}\tilde{I}_{0,1,1,1,1,1,1,0,0} \, ,\\[-0.3cm]
f_{47}&=\frac{\epsilon ^4}{\mu}( \tilde{I}_{0,1,1,1,1,1,1,-1,0}-\tilde{I}_{0,1,1,1,1,0,1,0,0}) \, ,\\[-0.3cm]
f_{48}&=\frac{\epsilon ^4 \sqrt{\mu+\nu} \sqrt{\mu^2+\mu+\nu}}{\mu^2 \nu} \tilde{I}_{1,0,1,1,1,1,1,0,0} \, ,\\[-0.3cm]
f_{49}&=\frac{\epsilon ^4}{\mu^3}(4 \tilde{I}_{1,0,1,1,1,1,1,0,0}+\mu \tilde{I}_{1,0,1,1,1,0,1,0,0}- \mu \tilde{I}_{1,0,1,1,1,1,1,-1,0})\, ,
\\[-0.3cm]
f_{50}&=\frac{\epsilon ^4 \sqrt{\mu+\nu+1} }{\mu^2 \nu}\left(\mu \tilde{I}_{1,1,1,1,1,1,1,-1,0}-4 \tilde{I}_{1,1,1,1,1,1,1,0,0}\right) + \cdots \, ,\\[-0.3cm]
f_{51}&=\frac{\epsilon ^4  \sqrt{\mu+\nu} \sqrt{\mu^2+\mu+\nu}}{\mu^2 \nu}\tilde{I}_{1,1,1,1,1,1,1,-1,0} + \cdots \, , \\[-0.3cm]
f_{52}&=\frac{\epsilon ^4 }{\mu^2 \nu}\left(\mu \nu \tilde{I}_{1,1,1,1,1,1,1,-1,-1}-2 \mu \tilde{I}_{1,1,1,1,1,1,1,-1,0}-2 \nu \tilde{I}_{1,1,1,1,1,1,1,0,-1}+8 \tilde{I}_{1,1,1,1,1,1,1,0,0}\right) + \cdots \, , \\[-0.3cm]
f_{53}&=\frac{ \epsilon ^4\sqrt{\mu+1}  }{\mu^2 \nu}\left(\mu \tilde{I}_{1,1,1,1,1,1,1,-1,0}+\nu \tilde{I}_{1,1,1,1,1,1,1,0,-1}-4 \tilde{I}_{1,1,1,1,1,1,1,0,0}\right) + \cdots \, , \\[-0.3cm]
f_{54}&=\frac{\epsilon ^4 }{\mu^2}\left(\mu \tilde{I}_{1,1,1,1,1,1,1,-2,0}-4 \tilde{I}_{1,1,1,1,1,1,1,-1,0}\right) + \cdots \, ,
\end{align*}
where each $\tilde{I}_{\{a_i\}}$ should be multiplied by a factor of $m_t^2$ to the power of $a-4$ (with $a \equiv \sum a_i$), such that all the $f_i$'s are dimensionless. Note that for $f_{50}$--$f_{54}$ we have only shown the integrals in the highest topology (with 7 propagators). Their dependencies on integrals in the sub-topologies are rather lengthy (denoted by the ellipses), but are easy to be recovered from their differential equations and the expressions for $f_{1}$--$f_{49}$.

\end{document}